\documentclass[a4]{article}    

\usepackage[english]{babel}
\usepackage{graphicx}
\usepackage{mathtools}
\usepackage{amsmath}
\usepackage{amssymb}
\usepackage{breakcites}
\usepackage{comment}
\usepackage{float}
\usepackage[margin=0.7in]{geometry}

\usepackage{color, colortbl}

\usepackage{amsthm}
\usepackage{hyperref}
\usepackage{subfigure}
\usepackage{tikz}
\usepackage{algorithm}

\usepackage{algorithmic}
\usepackage{tabularx}

\usetikzlibrary{positioning,chains,fit,shapes,calc}
\usetikzlibrary{backgrounds}
\usetikzlibrary{arrows.meta}
\usetikzlibrary{shapes,snakes}

\newtheorem{theorem}{Theorem}

\newtheorem{definition}[theorem]{Definition}
\newtheorem{corollary}[theorem]{Corollary}
\newtheorem{remark}[theorem]{Remark}
\definecolor{ao}{rgb}{0.55, 0.71, 0.0}
\definecolor{bleudefrance}{rgb}{0.19, 0.55, 0.91}
\definecolor{dimgray}{rgb}{0.41, 0.41, 0.41}    
\definecolor{mediumorchid}{rgb}{0.73, 0.33, 0.83}
\definecolor{mediumtealblue}{rgb}{0.0, 0.33, 0.71}
\definecolor{harvestgold}{rgb}{0.85, 0.57, 0.0}
\definecolor{blue(pigment)}{rgb}{0.2, 0.2, 0.6}
\definecolor{forestgreen(traditional)}{rgb}{0.27, 0.35, 0.27}
\definecolor{cadmiumred}{rgb}{0.89, 0.0, 0.13}
\definecolor{orange(webcolor)}{rgb}{1.0, 0.5, 0.0}
\usepackage{enumitem} 
\setlist[itemize]{noitemsep}

\title{Minimum Structural Sensor Placement for Switched Linear Time-Invariant Systems and Unknown Inputs}
\date{\today}

\author{$\text{Emily A. Reed}^{*}$,
$\text{Guilherme Ramos}$\footnote{Both authors contributed equally.
{\scriptsize E. Reed is with the Ming Hsieh Electrical and Computer Engineering Department at the University of Southern California, USA. G. Ramos is with the Department of Electrical and Computer Engineering, Faculty of Engineering, University of Porto, Portugal. P. Bogdan is a faculty member at the University of Southern California in the Ming Hsieh Electrical and Computer Engineering Department. S. Pequito is a faculty member at the Delft University of Technology in the Delft Center for Systems and Control.
 This work was supported in part by FCT project POCI-01-0145-FEDER-031411-HARMONY, National Science Foundation GRFP DGE-1842487, Career Award CPS/CNS-1453860, CCF-1837131, MCB-1936775, CNS-1932620, CMMI-1936624, CMMI 1936578, the University of Southern California Annenberg Fellowship, USC WiSE Top-Off Fellowship, the DARPA Young Faculty Award and DARPA Director Award N66001-17-1-4044. The views, opinions, and/or findings contained in this article are those of the authors and should not be interpreted as representing the official views or policies, either expressed or implied by the Defense Advanced Research Projects Agency, the Department of Defense or the National Science Foundation.}}, Paul Bogdan, Sérgio Pequito}
\begin{document}
\maketitle

\begin{abstract}                          
In this paper, we study the structural state and input observability of continuous-time switched linear time-invariant systems and unknown inputs. 
First, we provide necessary and sufficient conditions for their structural state and input observability that can be efficiently verified in $O((m(n+p))^2)$, where $n$ is the number of state variables, $p$ is the number of unknown inputs, and $m$ is the number of modes. 
Moreover, we address the minimum sensor placement problem for these systems by  adopting a feed-forward analysis and by providing an algorithm with a computational complexity of $ O((m(n+p)+\alpha)^{2.373})$, where $\alpha$ is the number of target  strongly connected components of the system's digraph representation. 
Lastly, we explore different assumptions on both the system and unknown inputs (latent space) dynamics that add more structure to the problem, and thereby, enable us to render algorithms with lower computational complexity, which are suitable for implementation in large-scale systems. 
\end{abstract}


\section{Introduction}\label{sec:introduction} 
Scientists and engineers model systems by describing the nature of their dynamics and the environment in which they interact. One powerful tool to model complex switching dynamics is to adopt a switched linear \mbox{time-invariant} framework. This model assumes that the system under scrutiny transitions between different (yet known) linear \mbox{time-invariant} dynamics, where such transitions are discrete in nature and are captured by a switching signal for which the sequence of the switches may not be known \emph{a priori}. Examples of such systems include the power electric grid \cite{du2015fault}, where the change in dynamics may be dictated by a faulty transmission line~\cite{ramos2013model,ramos2015analysis}, or a multi-agent system \cite{sun2017stabilization,ramos2020ijoc}, where the dynamics may change due to a loss in communication among agents.  

However, modeling scenarios often neglect the fact that the interaction of a dynamical system with its environment introduces errors. We can describe these external environmental errors by unknown inputs entering into the dynamical system. For instance, in the power grid, the generated power and/or the customer demand behave as unknown inputs. Similarly, in multi-agent robotic systems, particularly in surface vehicles, friction behaves as an unknown input, whereas in the context of unmanned aerial vehicles, airflow or ocean currents act as unknown inputs. An alternate scenario is in networked systems where the unknown input is due to the interconnections with the remaining hidden network~\cite{hutchison2013dynamic,alur2015principles,corradini2017sliding,xie2018secure,farivar2019artificial,gupta2018re}.  
As is evident in the previously mentioned examples, in control engineering, a recurrent practice is that of modeling the unknown inputs in a latent space that can capture the main features of the incoming signal but does not model the system from which the unknown input originates.   

To monitor such switched linear time-invariant systems under unknown inputs requires us to assess both the state and the inputs by guaranteeing that the system is state and input observable \cite{sundaram2012structural}. Often, however, we cannot accurately know the parameters of the system. 
Moreover, if the parameters are known, the study of controllability and/or observability properties leads to NP-hard problems~\cite{ramos2018robust}.   
Hence, we assume that only the structure of the system is known meaning that a system parameter is either zero or could take on any real scalar value \cite{ramos2020structural}. In this context, we can rely on the notion of structural state and input observability that yields state and input observability for almost all system parameterizations. 

Previous work has provided the necessary and sufficient conditions to ensure \emph{structural} state and input observability for discrete-time systems under unknown inputs \cite{sundaram2006designing}. Nonetheless, the counterpart for continuous-time switched linear-time invariant systems under unknown inputs were only studied in \cite{boukhobza2011observability} and \cite{boukhobza2011discrete,boukhobza2012sensor}. In particular,  \cite{boukhobza2011observability,boukhobza2012sensor} considers the graph-theoretic necessary and sufficient conditions for generic discrete mode observability of a continuous-time switched linear system with unknown inputs and proposed a computational method to verify such conditions with a complexity of $O(n^{6})$, where $n$ is the number of states. The works of \cite{boukhobza2011discrete,boukhobza2012sensor} present sufficient conditions for the generic observability of the discrete mode of continuous-time switched linear systems with unknown inputs and find an exhaustive location set to place sensors when these conditions are not satisfied with a computational complexity of $O(n^{4})$. However, none of these works considered the minimum number of required sensors and their placement to guarantee structural state and input observability as we consider in this work. This problem is important in designing control schemes for large scale systems and is often referred to as the \emph{minimum sensor placement}. While this problem has been studied for a variety of systems \cite{pequito2015framework}, to the best of the authors' knowledge, it has not been studied in the context of continuous-time switched linear time-invariant systems under unknown inputs.


The main contributions of this manuscript are as follows. We first provide necessary and sufficient conditions for structural state and input observability of continuous-time switched linear-time invariant systems under unknown inputs. Moreover, we can verify these conditions in $O((m(n+p))^2)$, where $n$ is the number of state variables, $p$ is the number of unknown inputs, and $m$ is the number of modes. Furthermore, we address the minimum sensor placement for these systems using a feed-forward analysis and an algorithm with a computational complexity of $O((m(n+p)+\alpha)^{2.373})$, where the $n\times n$ matrix multiplication algorithm with best asymptotic complexity runs in $ O(n^{\varsigma})$, with $\varsigma\approx 2.3728596$~\cite{alman2021refined}, and where $\alpha$ is the number of target strongly connected components of the system's digraph representation. We explore different assumptions on both the system and unknown input (latent space) dynamics to obtain more structure that enables us to provide new algorithms with lower computational complexity suitable to deal with large-scale systems. 
Finally, we present a real-world example from power systems to illustrate our results. 

We structure the remainder of our paper as follows. Section~\ref{sub:prob_stat} provides the addressed problem formulation. Section~\ref{sec:main_res} presents the main results including two graph-theoretic conditions for structural state and input observability for switched linear time-invariant systems with unknown inputs as well as an algorithm that determines the minimum set of state and input variables for ensuring structural state and input observability.
Section~\ref{sec:discussion} discusses several classes of switched linear time-invariant systems for which we can find a solution with a better computational complexity. 
Section~\ref{sec:real_world} provides a real-world example from power systems to illustrate our results. 
Finally, Section~\ref{sec:conclusions} concludes the paper and points out new directions for future research.

\section{Problem statement}\label{sub:prob_stat}
In this paper, we consider a continuous-time switched linear time-invariant (LTI) system with (unknown) inputs that can be described as follows:   
\begin{subequations}\label{eq:switched_lti}
\begin{align}
    \dot{x}(t) = & A_{\sigma(t)}x(t) + F_{\sigma(t)}d(t), \label{eq:switched_lti_a}\\
    \dot{d}(t) = & Q_{\sigma(t)}d(t), \label{eq:switched_lti_b}\\
    y(t) = & C_{\sigma(t)}x(t) + D_{\sigma(t)}d(t), \label{eq:switched_lti_c}
\end{align}
\end{subequations}
where $x(t)\in\mathbb{R}^{n}$ is the state, $d(t)\in\mathbb{R}^{p}$ represents the unknown inputs, $y(t)\in\mathbb{R}^{n}$ is the output, and $\sigma(t):[0,\infty)\rightarrow \mathbb M\equiv\{1,\dots,m\}$ is the unknown switching signal. System \eqref{eq:switched_lti} contains $m$ possible known subsystems also known as \emph{modes}, which we denote by the tuple $(A_{k},F_{k},Q_{k},C_k,D_k)$, where $\sigma(t) = k\in\mathbb M$.  Lastly, we implicitly assume that the dwell time of each mode is greater than 0.

In what follows, we seek to assess and determine the minimum sensor placement that ensures state and input observability for the continuous-time switched LTI system with unknown inputs in \eqref{eq:switched_lti}. 
\begin{definition}[State and Input Observability\\~\cite{molinari1976extended}]\label{def:observable}
     The switched LTI system described by ($A_{\sigma(t)}$, $F_{\sigma(t)}$, $Q_{\sigma(t)}$,  $C_{\sigma(t)}$, $D_{\sigma(t)}$, $\sigma(t)$; $T_f$) is said to be state and input observable for a time horizon $T_f$ \emph{if and only if} the initial state $x(t_0)$ and the unknown inputs $d(t)$ where $t\in [t_0,T_f]$ can be uniquely determined, given $(A_{\sigma(t)},F_{\sigma(t)},Q_{\sigma(t)},C_{\sigma(t)},D_{\sigma(t)}, \sigma(t); T_f)$ and measurements $y(t)$ $(t_0\leq t\leq T_f$).\hfill $\circ$
\end{definition}  

In this paper, we focus on the sensor placement problem. For the sake of simplicity, we assume that the measurements take the following form 
\begin{equation}\label{eq:switched_lti_cprime}
    y(t) = Cx(t) + Dd(t). \tag{1c'}
\end{equation}

Simply speaking, we assume that the output and feed-forward matrices are the same across all modes. Notice that this assumption can be waived as we discuss in the following Remark~\ref{wilog}.

{\centering
\allowdisplaybreaks
 \noindent \colorbox{gray!15}{\parbox{.935\columnwidth}{
\begin{remark}\label{wilog}
{\small We can consider a fixed set of measurements represented by $C$ and $D$ without loss of generality since taking the union of the measurements made in different modes, represented by $C_{\sigma(t)}$ and $D_{\sigma(t)}$, will result in the total set represented by $C$ and $D$.\hfill$\diamond$}
\end{remark}
}
}
}

We assume that each sensor is dedicated, meaning that each sensor can measure only one state or only one input. Considering an arbitrary set of sensors would lead to an NP-hard problem as this is the case for the linear time-invariant systems \cite{pequito2015framework}. We state this formally in the following assumption. 

\noindent\textbf{A1} The output matrix and feed-forward matrix are written as $C=\mathbb{I}_{n}^{\mathcal{J}_{x}}$ and $D=\mathbb{I}_{p}^{\mathcal{J}_{d}}$, where $\mathbb{I}_{n}^{\mathcal{J}_{x}}$ is a matrix where its rows are composed of canonical identity matrix rows that are each multiplied with any arbitrary value. These canonical rows are indexed by $\mathcal{J}_{x}=\{1,\dots,n\}$. Similarly, $\mathbb{I}_{p}^{\mathcal{J}_{d}}$ is a matrix where its rows are composed of canonical identity matrix rows that are each multiplied with any arbitrary value. These canonical vectors are indexed by $\mathcal{J}_{d}=\{1,\dots,p\}$.

Due to uncertainty in the system's parameters, we consider a structural systems framework \cite{ramos2020structural}. We introduce the following definition for a structural matrix. 
\begin{definition}\label{def:structural_matrix}
(Structural Matrix) A matrix $\bar M\in \{0,\star\}^{m_1\times m_2}$ is referred to as a structural matrix if $\bar M_{ij} = 0$, then $M_{ij}=0$, and if $\bar M_{ij}=\star$, then $M_{ij}\in\mathbb{R}$, so  $M_{ij}$ is any arbitrary real number and $M_{ij}$ is assumed to be independent of $M_{i'j'}$ for all $i,j,i',j'$ such that $i\neq i'$ and $j\neq j'$.
\end{definition}
With this notion in mind, we next define \emph{structural} state and input observability for the switched LTI system with unknown inputs in \eqref{eq:switched_lti}. 

\begin{definition}\label{def:struc_observable}
    (Structural State and Input Observability) The switched LTI system with unknown inputs described by the structural matrices \\$(\bar A_{\sigma(t)},\bar F_{\sigma(t)}, \bar Q_{\sigma(t)}, \bar C_{\sigma(t)},\bar D_{\sigma(t)}, \sigma(t); T_f)$ is said to be structurally state and input observable for a time horizon $T_f$ \emph{if and only if} there exists a system described by $(A_{\sigma(t)},F_{\sigma(t)},Q_{\sigma(t)},C_{\sigma(t)},D_{\sigma(t)}, \sigma(t); T_f)$ that is state and
input observable and satisfies the structural pattern imposed by the structural matrices $(\bar A_{\sigma(t)},\bar F_{\sigma(t)},\bar Q_{\sigma(t)},\bar C_{\sigma(t)},\bar D_{\sigma(t)})$.\hfill $\circ$
\end{definition}

Subsequently, the problem statement we seek to address in this paper is as follows: given $\bar A_{\sigma(t)}$, $\bar F_{\sigma(t)}$, $\bar Q_{\sigma(t)}$, which are the known structural matrices of system in \eqref{eq:switched_lti_a} and \eqref{eq:switched_lti_b}, and time horizon $T_f$, we aim to find the minimum set of states $\mathcal{J}_{x}$ and inputs $\mathcal{J}_{d}$ that need to be measured to ensure structural state and input observability. We present this formally as 
\begin{equation}\label{eq:problem1}\tag{$\mathcal P_1$}
    \begin{split}
        \min_{\substack{\mathcal{J}_{x}\subseteq \{1,\dots, n\}\\\mathcal{J}_{d}\subseteq \{1,\dots, p\}}} & \qquad |\mathcal{J}_{x}|+|\mathcal{J}_{d}|  \\
        \text{s.t.  } & (\bar A_{\sigma(t)},\bar F_{\sigma(t)},\bar Q_{\sigma(t)}, \mathbb{\bar I}_n^{\mathcal{J}_{x}},\mathbb{\bar I}_p^{\mathcal{J}_{d}}, \sigma(t); T_f)\\
        &\text{ is struct. state and input observable.}
    \end{split}
\end{equation}

For the sake of clarity, we assume that the matrix $\bar F_{\sigma(t)}$ does not have zero columns as this would correspond to having disturbances that do not affect the dynamics of the system. 

\section{Minimum Structural Sensor Placement for Switched LTI Systems with Unknown Inputs}\label{sec:main_res}

In this section, we proceed as follows. 
First, we provide necessary and sufficient conditions for the feasibility of the optimization problem $\mathcal P_1$. 
Second, we characterize the minimal solution of problem $\mathcal P_1$. Next, we develop an algorithm to obtain a solution to $\mathcal P_1$, and we assess its computational complexity. 
Lastly, we provide a discussion about the trade-offs between the assumptions on the dynamics and the algorithms used to solve the proposed problem. 

We start by introducing the notion of \emph{generic rank}, which allows us to provide conditions for structural state and input observability of continuous-time switched LTI systems with unknown inputs.


\begin{definition}\label{def:generic_rank}
    (Generic rank): The generic rank (g-rank) of an $n_1\times n_2$ structural matrix $\bar M$ is 
    \[
        g-rank(\bar M)=\max_{M\in [\bar M]}rank(M),
    \] 
    where $[\bar M] = \{M\in\mathbb R^{n_1\times n_2}: M_{i,j}=0 \text{ if } \bar M_{ij}=0\}$.\hfill$\circ$
\end{definition}

Next, we introduce several graph-theoretical and algebraic definitions required for defining the conditions for state and input observability of switched LTI systems with unknown inputs. 

A \emph{directed graph} associated with any structural system matrix $\bar M$ is constructed in the following manner. 
A directed graph is written as $\mathcal G({\bar M})=(\mathcal V,\mathcal E)$, where $\mathcal V$ denotes the set of vertices (or nodes) such that $\mathcal{V} = \mathcal{M}_x$, and $\mathcal E$ denotes the (directed) edges between the vertices in the graph such that $\mathcal{E}=\mathcal{E}_{\mathcal{M}_x,\mathcal{M}_x} = \{(m_{j},m_{i}):\bar M(i,j)\neq 0\}$. 
For a specific time $t'$ such that $\sigma(t')=k$, we associate the system in \eqref{eq:switched_lti_a} and \eqref{eq:switched_lti_b} with a system digraph $\mathcal{G} \equiv \mathcal{G}(\bar A_{k}, \bar F_{k}, \bar Q_{k}, \mathbb{I}_{n}^{\mathcal{J}_{x}}, \mathbb{I}_{p}^{\mathcal{J}_{d}})= (\mathcal{V}, \mathcal{E}^{k})$, where $\mathcal{V} = \mathcal{X}\cup\mathcal{D}\cup\mathcal{Y}$,  $\mathcal{X} = \{x_{1},\dots,x_{n}\}$, $\mathcal{D}=\{d_{1},\dots,d_{p}\}$, and $\mathcal{Y}=\{y_{1},\dots,y_{n}\}$ are the state, unknown input, and output vertices, respectively.
Furthermore, we have that $\mathcal{E}^{k}=\mathcal{E}^{k}_{\mathcal{X},\mathcal{X}}\cup\mathcal{E}^{k}_{\mathcal{D},\mathcal{X}}\cup\mathcal{E}^{k}_{\mathcal{D},\mathcal{D}}\cup\mathcal{E}_{\mathcal{X},\mathcal{Y}}\cup\mathcal{E}_{\mathcal{D},\mathcal{Y}}$, where $\mathcal{E}^{k}_{\mathcal{X},\mathcal{X}} = \{(x_{j},x_{i}): \bar A_{k}(i,j) \neq 0\}$, $\mathcal{E}^{k}_{\mathcal{D},\mathcal{X}} = \{(d_{j},x_{i}): \bar F_{k} \neq 0\}$, $\mathcal{E}^{k}_{\mathcal{D},\mathcal{D}}=\{(d_j,d_i): \bar Q_{k} \neq 0\}$, $\mathcal{E}_{\mathcal{X},\mathcal{Y}} = \{(x_{j},y_{i}): \mathbb{I}_{n}^{\mathcal{J}_x}(i,j) \neq 0\}$, and $\mathcal{E}_{\mathcal{D},\mathcal{Y}} = \{(d_{j},y_{i}):
\mathbb{I}_{p}^{\mathcal{J}_d}(i,j) \neq 0\}$ are the state, input, and output edges, respectively.

Next, we introduce a mathematical operator, which plays a key role in presenting the conditions for structural state and input observability of switched LTI systems with unknown inputs. 
\begin{definition} (Union of structural matrices)
The mathematical operator $\vee$ is an entry-wise operation such that a structural matrix $\bar A = \bigvee_{k=1}^m\bar{ A_k} = \bar A_{1}\vee \bar A_{2} \vee \dots \vee \bar A_{m}$ has a non-zero entry at $(i,j)$ if at least one of the matrices $\bar A_{k}$ has a non-zero entry in that same location $(i,j)$, and $\bar A(i,j) = 0$, otherwise.\hfill$\circ$ 
\end{definition}


With this definition, we introduce the directed graphs $\mathcal{G}\left(\bigvee_{k=1}^{m}\bar A_{k}'\right)$ and $\mathcal{G}\left(\bigvee_{k=1}^{m}\bar A_{k}', \bar C'\right)$. More specifically,  $\mathcal{G}\left(\bigvee_{k=1}^{m}\bar A_{k}'\right)=(\mathcal{X}', \mathcal{E}_{\mathcal{X}',\mathcal{X}'})$ where $\mathcal{E}_{\mathcal{X}'}=\{(x_j',x_i'): \bigvee_{k=1}^{m} \bar A_{k_{i,j}}'\neq 0\}$. In addition,  $\mathcal{G}\left(\bigvee_{k=1}^{m}\bar A_{k}', \bar C'\right)=(\mathcal{V}, \mathcal{E})$ where $\mathcal{V}=\mathcal{X}'\cup\mathcal{Y}'$ and $\mathcal{E}=\mathcal{E}_{\mathcal{X}',\mathcal{X}'} \cup \mathcal{E}_{\mathcal{X}',\mathcal{Y}'}$ such that $\mathcal{E}_{\mathcal{X}',\mathcal{X}'} = \{(x_j',x_i'): \bigvee_{k=1}^{m} \bar A_{k_{i,j}}'\neq 0\}$ and $\mathcal{E}_{\mathcal{X}',\mathcal{Y}'} = \{(x_j', y_i'): C'_{i,j} \neq 0\}$. We next introduce the necessary and sufficient conditions for structural state and input observability for continuous-time switched LTI systems with unknown inputs. 

\begin{theorem}[Necessary and sufficient conditions for structural state and input observability]\label{th:struct_obs}
A continuous-time switched LTI system with unknown inputs in~\eqref{eq:switched_lti_a}, \eqref{eq:switched_lti_b}, and \eqref{eq:switched_lti_cprime} is structurally state and input observable \underline{if and only if} the next two conditions hold: 
\begin{enumerate}
    \item[(i)] $\mathcal G\left(\displaystyle\bigvee_{k=1}^m\bar{ A_k'},
    \bar{C}'\right)$ has all state vertices that access at least one output vertex; 
    \item[(ii)] $\text{g-rank}\left([\bar A_1';\ldots;\bar A_m';\bar {C'}
    ]\right)=n+p$, 
\end{enumerate}
where, for $k\in\mathbb M$ and the matrices $\bar A_k'$ and $\bar C'$ are defined as 
$\bar A_k'= 
\left[\begin{smallmatrix}
     \bar Q_k   &  0   \\
     \bar F_k &  \bar A_k 
\end{smallmatrix}\right]
\,\text{ and }\,
{\bar C}' = 
\left[\begin{smallmatrix}
     \bar D & \bar C 
\end{smallmatrix}\right].$\hfill$\circ$
\end{theorem}

{\centering
\allowdisplaybreaks
\noindent \colorbox{gray!15}{\parbox{.935\columnwidth}{
\begin{remark}
{\small 
Consider a switching signal that ensures the structural observability of the switched linear continuous-time systems.  
The order of transitions of system modes does not influence its structural observability. 
This property comes from the fact that: 
\begin{itemize}
    \item the `` $\vee$'' operation, in condition (i) Theorem~\ref{th:struct_obs}, is commutative;
    \item a permutation of the matrices, in condition (ii) Theorem~\ref{th:struct_obs},  yields the same $\text{g-rank}$.~\hfill $\diamond$ 
\end{itemize}
}
\end{remark}
}
}
}

Next, we define a few other important graph-theoretic concepts. A \emph{bipartite graph} denoted as $\mathcal{B}$ associates a matrix $M$ of dimension $n_1\times n_2$ to two vertex sets $\mathcal V_{r}=\{1,\dots,n_1\}$ and $\mathcal V_{c}=\{1,\dots,n_2\}$, which are the set of row and column vertices, respectively. The connections in the matrix $M$ relate to the connections between vertex sets $\mathcal V_{r}$ and $\mathcal V_{c}$ by an edge set $\mathcal{E}_{\mathcal V_{c},\mathcal V_{r}} = \{(v_{c_{j}},v_{r_{i}}):M_{ij} \neq 0\}$ thereby allowing the bipartite graph of matrix $M$ to be written as $\mathcal{B}(\mathcal V_{c},\mathcal V_r,\mathcal{E}_{\mathcal V_{c},\mathcal V_{r}})$. 
A \emph{matching} is a collection of edges where the beginning vertex is different from the ending vertex for all edges in the set and there are no two edges in the set that have any of the same vertices. A \emph{maximum matching} is the matching that has the maximum number of edges among all possible matchings. A \emph{weighted bipartite graph} of a matrix $M$, denoted as $\mathcal{B}((\mathcal V_c,\mathcal V_r,\mathcal{E}_{\mathcal V_{c},\mathcal V_{r}}),w)$, has weights $w: \mathcal{E}_{\mathcal V_{c},\mathcal V_{r}}\rightarrow\mathbb{R}$ associated with the edges in the bipartite graph. Finding the maximum matching such that the sum of the weights is minimized in the weighted bipartite graph is called the \emph{minimum weight maximum matching} (MWMM). 


Now, we must introduce the notions of a strongly connected component and non-accessible states. 
Let $\mathbb Z_{\geq 0}$ denote the set of non-negative integers. 
First, we define a \emph{path} of size $l\in\mathbb Z_{\geq 0}$ as a sequence of vertices, $p_s=(v_1,v_2,\ldots,v_l)$, where the vertices do not repeat, $v_i\neq v_j$ for $i\neq j$, and $(v_i,v_{i+1})$ is an edge of the directed graph for $i=1,\ldots l-1$. A \emph{subgraph} denoted by $\mathcal{G}(\mathcal{V}',\mathcal{E}')$ is a subset of vertices $\mathcal{V}'\subset \mathcal{V}$ and its corresponding edges $\mathcal{E}'\subset\mathcal{E}$ of a particular graph $\mathcal{G}(\mathcal{V},\mathcal{ E})$. A \emph{connected component} is any subgraph with paths that connect any two vertices in the subgraph. A connected component is said to be a \emph{strongly connected component} (SCC) if the subgraph is maximal meaning there is no other subgraph that contains the maximal subgraph. 
A \emph{sink SCC} is a strongly connected component that is connected to an output vertex. A \emph{source SCC} is a strongly connected component that is connected to an input vertex. A \emph{target-SCC} is a strongly connected component that does not have any outgoing edges. We note that every digraph can be represented as a directed acyclic graph (DAG), where each node in the DAG represents an SCC in the digraph. 
Finally, a \emph{non-accessible state} is one that does not have a path to an output vertex (either measuring a state or input). 

We present graph-theoretic conditions for structural state and input observability of continuous-time switched LTI systems with unknown inputs. 


\begin{corollary}\label{th:struct_obs_sol}
A switched LTI continuous-time system~\eqref{eq:switched_lti} is structurally observable \underline{if and only if} the next two conditions hold: 
\begin{enumerate}
    \item[(i)] there exists an edge from one state variable of each target-SCC of $\mathcal G\left(\bigvee_{k=1}^m\bar A_k'\right)$ to an output variable of $\mathcal G\left(\bigvee_{k=1}^m\bar A_k',
    \bar C'\right)$; 
    \item[(ii)] 
    $\mathcal B\left( [\bar A_1'; \ldots; \bar A_m';
    \bar C' ]\right)$ 
    has a maximum matching of size~$n+p$; 
\end{enumerate}
where, for $k\in\mathbb M$, the matrices $\bar A_k'$ and $\bar C'$ are defined as 
$\bar A_k'= 
\text{\small $\left[\begin{smallmatrix}
    \bar Q_k   &  \mathbf{0}   \\
     \bar F_k & \bar A_k 
\end{smallmatrix}\right]$}
\,\text{ and }\,
{\bar C}' = 
\text{\small $\left[\begin{smallmatrix}
     \bar D & \bar C 
\end{smallmatrix}\right]$}.$
\hfill$\circ$
\end{corollary}

In the following remark, we outline the computational complexity in which we can verify the conditions of Corollary~\ref{th:struct_obs_sol}. 

{\centering
\allowdisplaybreaks
\noindent \colorbox{gray!15}{\parbox{.935\columnwidth}{
\begin{remark}
{\small
We can verify the two conditions in Corollary~\ref{th:struct_obs_sol} in $O((m(n+p))^2)$, where $n$ is the number of state variables, $p$ is the number of unknown inputs, and $m$ is the number of modes (Section 3.3, \cite{liu2013structural}). We notice that the number of variables required to be measured is always less than or equal to $n+p$.\hfill$\diamond$
}
\end{remark}
}
}
}

With the graph-theoretic conditions for structural state and input observability enumerated, we introduce Algorithm \ref{alg:main}. Briefly, the algorithm finds the minimum set of state and input variables to ensure that the conditions of Corollary~\ref{th:struct_obs_sol} are satisfied. First, the algorithm finds the maximum collection of variables that satisfy the condition of Corollary~\ref{th:struct_obs_sol} by constructing the MWMM of $\mathcal{B}([\bar A_1';\ldots; \bar A_m';\bar T])$, where $\bar T$ has as many rows as target-SCCs, and the non-zero column entries of $\bar T$ specify the indices of the  augmented states that make up each target-SCC. Furthermore, weights are considered on the edges of the bipartite graph such that all edge weights are zero unless the edges connect to a vertex established by $\bar T$ at which the weight is set to one. If there is an edge in the MWMM that has a weight of one, then the index of the column vertex connecting the edge is the same index of the augmented state variable that satisfies both conditions in Corollary~\ref{th:struct_obs_sol}. The algorithm then proceeds to find the minimum set of variables from the maximum collection that still ensure the conditions of Corollary~\ref{th:struct_obs_sol}. 

\begin{algorithm}[H]
		{\small
		\caption{Dedicated solution to $\mathcal P_1$}
		\label{alg:main}
		\begin{algorithmic}[1]
			\STATE{\textbf{Input}: A structural switched LTI system with $\mathbb M=\{1,\ldots,m\}$ modes described by $\{\bar A_1,\ldots,\bar A_m,\bar F_1,\hdots,\bar F_m,\bar Q_1,\hdots,\bar Q_m\}$, where $\bar A_k\in\{0,\star\}^{n\times n}$, $\bar F_k\in\{0,\star\}^{n\times p}$, $\bar Q_k\in\{0,\star\}^{p\times p}, \forall k\in\{1,\ldots,m\}$}
			\STATE{\textbf{Output}: Output $\bar C = \mathbb{\bar I}_n^{\mathcal{J}_{x}}$ and $\bar D= \mathbb{\bar I}_p^{\mathcal{J}_{d}}$, where $\mathcal{J} = \mathcal{J}_{x} \cup \mathcal{J}_{d}$, $\mathcal{J}_{d}=\{i\in\mathcal J:i \leq p\}$, and $\mathcal{J}_{x}=\{i\in\mathcal J:i> p\}$  }
            \STATE{\textbf{Set} $\bar A_k'= 
\left[\begin{smallmatrix}
     \bar Q_k   &  \mathbf{0}   \\
     \bar F_k &  \bar A_k 
\end{smallmatrix}\right]
$}

			\STATE{\textbf{Compute} the $\alpha$ target-SCCs of $\mathcal G\left(\bigvee_{k=1}^m\bar A_k'\right)=(\mathcal X',\mathcal E_{\mathcal X',\mathcal X}')$, denoted by $\{\mathcal S_1,\ldots,\mathcal S_\alpha\}$}
			\STATE{\textbf{Build} the bipartite graph $\mathcal B([\bar A_1'; \ldots; \bar A_m'; \bar T])=(\mathcal V_c,\mathcal V_r,\mathcal E_{\mathcal V_c,\mathcal V_r})$, where $\bar T\in\{0,\star\}^{(n+p)\times\alpha}$ and $\bar T_{i,j}=\star$ if ${x'}_j\in \mathcal S_i$ and $\bar T_{i,j}=0$, otherwise. 
			We denote the rows of matrix $\bar A_k'$ by $\{r_1^k,\ldots,r_{n+p}^k\}$, and the rows of $\bar T$ by $\{t_1,\ldots,t_\alpha\}$.} 
			\STATE{\textbf{Set} the weight of the edges $e\in\mathcal E_{\mathcal V_c,\mathcal V_r}$ to 
			$
			\text{\small $\begin{cases}
			1, & \text{if }e\in (\{t_1,\ldots,t_\alpha\}\times\mathcal V_c)\cap \mathcal E_{\mathcal V_c,\mathcal V_r}\\
			0, &\text{otherwise}
			\end{cases}$}
			.$}
			\STATE{\textbf{Find} a MWMM $\mathcal M'$ of the bipartite graph computed in Step 5, with edges' costs of Step. 6.}
			\STATE{\textbf{Set}} the column vertices associated with $\bar T$ belonging to $\mathcal M'$, i.e., $\mathcal J'=\{i\,:\,(t_j,c_i)\in\mathcal M', j\in\{1,\ldots,\alpha\}\text{ and }c_i\in\mathcal V_c\}$ and 
			$\mathcal T=\{j\,:\,(t_j,c_i)\in\mathcal M', j\in\{1,\ldots,\alpha\}\}$
			\STATE{\textbf{Set} $\mathcal J''=\{1,\ldots,n+p\}\setminus\{i\in\{1,\ldots,n+p\}\,:\,(r^k_j,c_i)\in\mathcal M',k\in\{1,\ldots,m\},j\in\{1,\ldots,n+p\}\} $}  
			\STATE{\textbf{Set} $\mathcal J'''$ to contain one and only one index of a state variable from each target-SCC in $\{\mathcal S_s\,:\,s\in\{t_1,\ldots,t_\alpha\}\setminus \mathcal T\}$}  
			\STATE{\textbf{Set} $\mathcal J=\mathcal J'\cup\mathcal J''\cup\mathcal J'''$} 
			\STATE{\textbf{Set} $\mathcal{J}_{d}=\{i\in\mathcal J:i \leq p\}$, and $\mathcal{J}_{x}=\{i\in\mathcal J:i> p\}$ }
		\end{algorithmic} 
		}
\end{algorithm} 

In the next result, we show that Algorithm \ref{alg:main} finds the minimum set of states and inputs to ensure structural state and input observability. 
\begin{theorem}\label{th:soundness_and_complexity}
Algorithm~\ref{alg:main} is sound, i.e., it provides a solution to $\mathcal P_1$, and the computational complexity of Algorithm~\ref{alg:main} is $O((m(n+p)+\alpha)^\varsigma)$, where $\varsigma<2.373$ is the exponent of the best known computational complexity of performing the product of two square matrices~\cite{alman2021refined}.~\hfill$\circ$
\end{theorem}

{\centering
\allowdisplaybreaks
\noindent \colorbox{gray!15}{\parbox{.935\columnwidth}{
\begin{remark}
{\small The computational complexity presented in Theorem~\ref{th:soundness_and_complexity} might not be amenable for ensuring the sensor placement for very large systems. Nonetheless, there are some particular classes of systems for which algorithms with lower computational complexity can be devised. In the next section, we present these classes of systems.\hfill$\diamond$}
\end{remark}
}
}
}

\subsection*{Example 1}
Let us consider the following linear continuous-time system with two modes $\{\bar{A}_{k}, \bar{F}_{k},\bar{Q}_{k}\}_{k=1}^{2}$, where $\bar{A}_{k}\in\{0,1\}^{5\times 5}$, $\bar{F}_{k}\in\{0,1\}^{5\times 1}$, $\bar Q_k\in \{0,1\}$. In particular, $[\bar A_{1}]_{3,1} = 1$, $[\bar A_{1}]_{2,2} = 1$, $[\bar A_{1}]_{4,3} = 1$, $[\bar A_{2}]_{3,2} = 1$, $[\bar A_{2}]_{5,3} = 1$, $[\bar F_{1}]_{2,1} = 1$, $[\bar F_{2}]_{2,1} = 1$, and $\bar Q_k=0,\forall k\in\{1,2\}$.

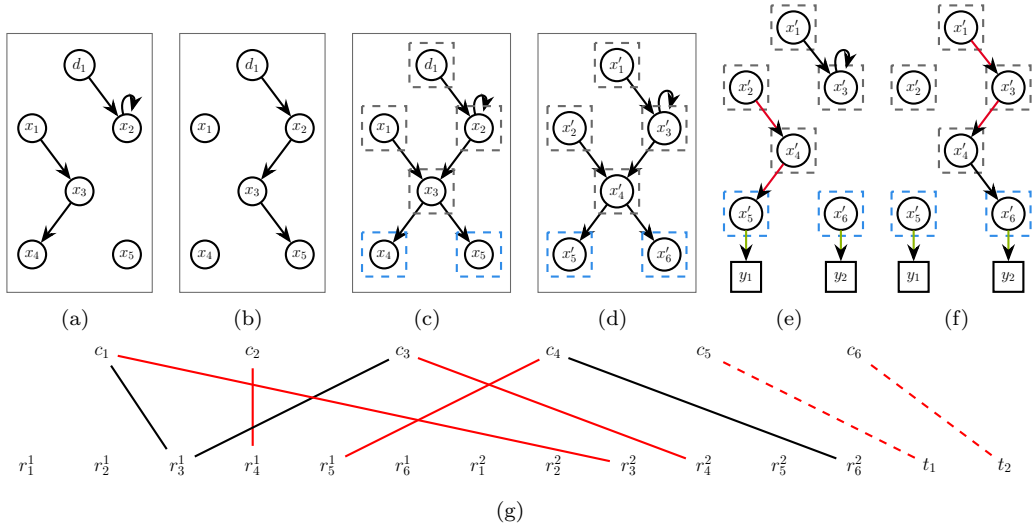
\begin{figure}[H]
\centering

\subfigure[]{
\begin{tikzpicture}[scale=.42, transform shape,node distance=1.5cm]
\begin{scope}[every node/.style={circle,thick,draw},square/.style={regular polygon,regular polygon sides=4},none/.style={regular polygon,regular polygon sides=1}]
    \node (x1) at (1.5,0) {\Large $d_1$};
    \node (x2) at (0,-2) {\Large $x_1$};
    \node (x3) at (3,-2) {\Large $x_2$};
    \node (x4) at (1.5,-4) {\Large $x_3$};
    \node (x5) at (0,-6) {\Large $x_4$};
    \node (x6) at (3,-6) {\Large $x_5$};    
\end{scope}

\begin{scope}[>={Stealth[black]},
              every edge/.style={draw=black, thick},eblue/.style={draw=bleudefrance, thick}]
      
    \path [->] (x1) edge node {} (x3);
    \path [->] (x2) edge node {} (x4);
    \path [->] (x4) edge node {} (x5);
    \path [->] (x3) edge[loop above] node {} (x3);

\end{scope}
 \draw [draw=dimgray] (-0.8,1) rectangle (3.8,-7.2);
\end{tikzpicture}
}
\subfigure[]{
\begin{tikzpicture}[scale=.42, transform shape,node distance=1.5cm]
\begin{scope}[every node/.style={circle,thick,draw},square/.style={regular polygon,regular polygon sides=4},none/.style={regular polygon,regular polygon sides=1}]
    \node (x1) at (1.5,0) {\Large $d_1$};
    \node (x2) at (0,-2) {\Large $x_1$};
    \node (x3) at (3,-2) {\Large $x_2$};
    \node (x4) at (1.5,-4) {\Large $x_3$};
    \node (x5) at (0,-6) {\Large $x_4$};
    \node (x6) at (3,-6) {\Large $x_5$}; 
    
\end{scope}

\begin{scope}[>={Stealth[black]},
              every edge/.style={draw=black, thick},eblue/.style={draw=bleudefrance, thick}]
      
    \path [->] (x1) edge node {} (x3);
    \path [->] (x3) edge node {} (x4);
    \path [->] (x4) edge node {} (x6);

\end{scope}

 \draw [draw=dimgray] (-0.8,1) rectangle (3.8,-7.2);
\end{tikzpicture}
}
\subfigure[]{
\begin{tikzpicture}[scale=.42, transform shape,node distance=1.5cm]
\begin{scope}[every node/.style={circle,thick,draw},square/.style={regular polygon,regular polygon sides=4},none/.style={regular polygon,regular polygon sides=1}]
    \node (x1) at (1.5,0) {\Large $d_1$};
    \node (x2) at (0,-2) {\Large $x_1$};
    \node (x3) at (3,-2) {\Large $x_2$};
    \node (x4) at (1.5,-4) {\Large $x_3$};
    \node (x5) at (0,-6) {\Large $x_4$};
    \node (x6) at (3,-6) {\Large $x_5$};  
\end{scope}

\begin{scope}[>={Stealth[black]},
              every edge/.style={draw=black, thick},eblue/.style={draw=bleudefrance, thick}]
      
    \path [->] (x1) edge node {} (x3);
    \path [->] (x3) edge node {} (x4);
    \path [->] (x4) edge node {} (x6);
    \path [->] (x2) edge node {} (x4);
    \path [->] (x4) edge node {} (x5);
    \path [->] (x3) edge[loop above] node {} (x3);

\end{scope}
\draw [draw=dimgray] (-1,1) rectangle (4,-7.2);
\draw [draw=dimgray,dashed,thick] (0.8,0.7) rectangle (2.2,-0.7);
\draw [draw=dimgray,dashed,thick] (-0.7,-1.3) rectangle (0.7,-2.7);
\draw [draw=dimgray,dashed,thick] (2.3,-1.3) rectangle (3.7,-2.7);
\draw [draw=dimgray,dashed,thick] (0.8,-3.3) rectangle (2.2,-4.7);
\draw [draw=bleudefrance,dashed,thick] (-0.7,-5.3) rectangle (0.7,-6.7);
\draw [draw=bleudefrance,dashed,thick] (2.3,-5.3) rectangle (3.7,-6.7);

\end{tikzpicture}
}
\subfigure[]{
\begin{tikzpicture}[scale=.42, transform shape,node distance=1.5cm]
\begin{scope}[every node/.style={circle,thick,draw},square/.style={regular polygon,regular polygon sides=4},none/.style={regular polygon,regular polygon sides=1}]
    \node (x1) at (1.5,0) {\Large $x_1'$};
    \node (x2) at (0,-2) {\Large $x_2'$};
    \node (x3) at (3,-2) {\Large $x_3'$};
    \node (x4) at (1.5,-4) {\Large $x_4'$};
    \node (x5) at (0,-6) {\Large $x_5'$};
    \node (x6) at (3,-6) {\Large $x_6'$};   
\end{scope}

\begin{scope}[>={Stealth[black]},
              every edge/.style={draw=black, thick},eblue/.style={draw=bleudefrance, thick}]
      
    \path [->] (x1) edge node {} (x3);
    \path [->] (x3) edge node {} (x4);
    \path [->] (x4) edge node {} (x6);
    \path [->] (x2) edge node {} (x4);
    \path [->] (x4) edge node {} (x5);
    \path [->] (x3) edge[loop above] node {} (x3);

\end{scope}
\draw [draw=dimgray] (-1,1) rectangle (4,-7.2);
\draw [draw=dimgray,dashed,thick] (0.8,0.7) rectangle (2.2,-0.7);
\draw [draw=dimgray,dashed,thick] (-0.7,-1.3) rectangle (0.7,-2.7);
 \draw [draw=dimgray,dashed,thick] (2.3,-1.3) rectangle (3.7,-2.7);
  \draw [draw=dimgray,dashed,thick] (0.8,-3.3) rectangle (2.2,-4.7);
  \draw [draw=bleudefrance,dashed,thick] (-0.7,-5.3) rectangle (0.7,-6.7);
  \draw [draw=bleudefrance,dashed,thick] (2.3,-5.3) rectangle (3.7,-6.7);
\end{tikzpicture}
}
\subfigure[]{
\begin{tikzpicture}[scale=.42, transform shape,node distance=1.5cm]
\draw [draw=dimgray,dashed,thick] (0.8,0.6) rectangle (2.2,-0.8);
\draw [draw=dimgray,dashed,thick] (-0.7,-1.3) rectangle (0.7,-2.7);
 \draw [draw=dimgray,dashed,thick] (2.3,-1.3) rectangle (3.7,-2.7);
  \draw [draw=dimgray,dashed,thick] (0.8,-3.3) rectangle (2.2,-4.7);
  \draw [draw=bleudefrance,dashed,thick] (-0.7,-5.3) rectangle (0.7,-6.7);
  \draw [draw=bleudefrance,dashed,thick] (2.3,-5.3) rectangle (3.7,-6.7);

\begin{scope}[every node/.style={circle,thick,draw},square/.style={regular polygon,regular polygon sides=4},none/.style={regular polygon,regular polygon sides=1}]
    \node (x1) at (1.5,-0.1) {\Large $x_1'$};
    \node (x2) at (0,-2) {\Large $x_2'$};
    \node (x3) at (3,-2) {\Large $x_3'$};
    \node (x4) at (1.5,-4) {\Large $x_4'$};
    \node (x5) at (0,-6) {\Large $x_5'$};
    \node (x6) at (3,-6) {\Large $x_6'$}; 
    
    \node[square,draw] (y1) at (0,-8) {\Large $y_1$};
    \node[square,draw] (y2) at (3,-8) {\Large $y_2$};
\end{scope}

\begin{scope}[>={Stealth[black]},
              every edge/.style={draw=black, thick},eblue/.style={draw=bleudefrance, thick}]
      
    \path [->] (x2) edge[cadmiumred] node {} (x4);
    \path [->] (x4) edge[cadmiumred] node {} (x5);
    
    \path [->] (x1) edge node {} (x3);
    \path [->] (x5) edge[ao] node {} (y1);
     \path [->] (x6) edge[ao] node {} (y2);
    \path [->] (x3) edge[loop above] node {} (x3);

\end{scope}
\end{tikzpicture}
}
\subfigure[]{
\centering
\begin{tikzpicture}[scale=.42, transform shape,node distance=1.5cm]
\draw [draw=dimgray,dashed,thick] (0.8,0.6) rectangle (2.2,-0.8);
\draw [draw=dimgray,dashed,thick] (-0.7,-1.3) rectangle (0.7,-2.7);
 \draw [draw=dimgray,dashed,thick] (2.3,-1.3) rectangle (3.7,-2.7);
  \draw [draw=dimgray,dashed,thick] (0.8,-3.3) rectangle (2.2,-4.7);
  \draw [draw=bleudefrance,dashed,thick] (-0.7,-5.3) rectangle (0.7,-6.7);
  \draw [draw=bleudefrance,dashed,thick] (2.3,-5.3) rectangle (3.7,-6.7);

\begin{scope}[every node/.style={circle,thick,draw},square/.style={regular polygon,regular polygon sides=4},none/.style={regular polygon,regular polygon sides=1}]
    \node (x1) at (1.5,-0.1) {\Large $x_1'$};
    \node (x2) at (0,-2) {\Large $x_2'$};
    \node (x3) at (3,-2) {\Large $x_3'$};
    \node (x4) at (1.5,-4) {\Large $x_4'$};
    \node (x5) at (0,-6) {\Large $x_5'$};
    \node (x6) at (3,-6) {\Large $x_6'$}; 
    
    \node[square,draw] (y1) at (0,-8) {\Large $y_1$};
    \node[square,draw] (y2) at (3,-8) {\Large $y_2$};
\end{scope}

\begin{scope}[>={Stealth[black]},
              every edge/.style={draw=black, thick},eblue/.style={draw=bleudefrance, thick}]
      
    \path [->] (x1) edge[cadmiumred] node {} (x3);
    \path [->] (x3) edge[cadmiumred] node {} (x4);
    \path [->] (x4) edge node {} (x6);
    
     \path [->] (x6) edge[ao] node {} (y2);
     \path [->] (x5) edge[ao] node {} (y1);

\end{scope}

\end{tikzpicture}
}

\vspace{-0.3cm}
\subfigure[]{
\begin{tikzpicture}[scale=.5, transform shape,node distance=1.5cm]
\begin{scope}[every node/.style={circle,draw=white},square/.style={regular polygon,regular polygon sides=4},none/.style={regular polygon,regular polygon sides=1}]
    \node (d1_1) at (0,0) {\Large $r_1^1$};
    \node (x1_1) at (2,0) {\Large $r_2^1$};
    \node (x2_1) at (4,0) {\Large $r_3^1$};
    \node (x3_1) at (6,0) {\Large $r_4^1$};
    \node (x4_1) at (8,0) {\Large $r_5^1$};
    \node (x5_1) at (10,0) {\Large $r_6^1$};
    \node (d1_2) at (12,0) {\Large $r_1^2$};
    \node (x1_2) at (14,0) {\Large $r_2^2$};
    \node (x2_2) at (16,0) {\Large $r_3^2$};
    \node (x3_2) at (18,0) {\Large $r_4^2$};
    \node (x4_2) at (20,0) {\Large $r_5^2$};
    \node (x5_2) at (22,0) {\Large $r_6^2$};
    
    \node (t1) at (24,0) {\Large $t_{1}$};
    \node (t2) at (26,0) {\Large $t_{2}$};
    
    
    \node (d1) at (2,3) {\Large $c_{1}$};
    \node (x1) at (6,3) {\Large $c_{2}$};
    \node (x2) at (10,3) {\Large $c_{3}$};
    \node (x3) at (14,3) {\Large $c_{4}$};
    \node (x4) at (18,3) {\Large $c_{5}$};
    \node (x5) at (22,3) {\Large $c_{6}$};
    
\end{scope}

\begin{scope}[>={Stealth[black]},
              every edge/.style={draw=black, thick},eblue/.style={draw=bleudefrance, thick}]
      
    \path [-] (d1) edge node {} (x2_1);
    \path [-] (x1) edge[red] node {} (x3_1);
    \path [-] (x3) edge[red] node {} (x4_1);
    \path [-] (x2) edge node {} (x2_1);
    \path [-] (d1) edge[red] node {} (x2_2);
    \path [-] (x2) edge[red] node {} (x3_2);
    \path [-] (x3) edge node {} (x5_2);

    \path [dashed] (t1) edge[red] node {} (x4);
    \path [dashed] (t2) edge[red] node {} (x5);
\end{scope}
\end{tikzpicture}
}
\vspace{-0.3cm}
\caption{(a) shows $A_{1}'$, (b) shows $A_{2}'$ (c) shows the union of the two modes of the continuous-time system with unknown inputs. (d) shows the augmented continuous-time system. Finally, (e) and (f) show the SCCs in dotted black rectangles, the target-SCC in a dotted blue rectangle, and the minimal output sensors and their placement for modes 1 and 2 respectively. (g) shows the bipartite graph $\mathcal{B}([\bar{A}_{1}'; \bar{A}_{2}';\bar T])$ with unitary weight on the dotted edge and zero weight on the solid edges. The collection of red edges is the minimum weight maximum matching.}
\label{fig:example_1}
\end{figure}

The individual modes of the system are shown in Figure \ref{fig:example_1} (a)-(b). We apply Algorithm \ref{alg:main} to this system to find the minimum set of dedicated sensors to achieve structural observability. We start by finding the union of the modes. $\bigvee_{k=1}^{m}\bar{A}_{k}'$ is given as follows
\begin{equation*}
    \bar{A}_{1}'\vee\bar{A}_{2}'= 
    \text{\small $\left[\begin{array}{@{}c|ccccc@{}}
    0 & 0 & 0 & 0 & 0 & 0 \\
    \hline
    0 & 0 & 0 & 0 & 0 & 0 \\
    1 & 0 & 1 & 0 & 0 & 0 \\
    0 & 1 & 1 & 0 & 0 & 0 \\
    0 & 0 & 0 & 1 & 0 & 0 \\
    0 & 0 & 0 & 1 & 0 & 0 
  \end{array}\right]
  $}
\end{equation*}
and is shown in Figure \ref{fig:example_1} (c) and (d), where (c) shows the union of the modes before the system has been augmented and (d) shows the union of the modes after the system has been augmented and relabeled with state $x'=[d^{\intercal}$  $x^{\intercal}]^{\intercal}$. 
With the system properly combined and augmented, we continue by finding the target-SCCs of $\mathcal{G}(\bigvee_{k=1}^{m}\bar{A}_{k}')$. We find that there are 6 SCCs, which are outlined in dashed rectangle boxes in Figures \ref{fig:example_1} (c) and (d). 
There are 2 target-SCCs, which are outlined in a dashed blue box in Figure \ref{fig:example_1} (c) and (d), so $\alpha=2$.

Next, we construct the bipartite graph $\mathcal{B}([\bar{A}_{1}';\bar{A}_{2}';\bar T])$. Since there are two target-SCCs each composed of one state ($x_{5}'$, $x_6'$), then $\bar T$ is as $
    \bar T = \left[\begin{smallmatrix}
    0 & 0 & 0 & 0 & 0 & 1\\
    0 & 0 & 0 & 0 & 1 & 0
    \end{smallmatrix}\right].$ 
In the bipartite graph, depicted in Figure \ref{fig:example_1} (g), there are two edges that each connect to one of the two target-SCCs and thus have unitary weight (shown as a dotted line) while the rest of the edges have zero weight (shown as solid lines). In step 7, we find the MWMM $\mathcal{M}'$, which is shown by the collection of red edges in Figure \ref{fig:example_1} (g). 
From the MWMM, we see that there are two edges in the MWMM that are connected to a target-SCC, so $\mathcal{J}' = \{5,6\}$. In step 9, we find $\mathcal{J}''$, which is the set of indices associated with the column vertices of $\bar A_k', \forall k$ that are not in the MWMM (i.e., all of the indices $i=\{1,\dots,n+p\}$ such that $c_{i}$ are not connected to a red edge in Figure \ref{fig:example_1} (g)). We notice that there are several different possible minimum weight maximum matchings. One of these possibilities is shown in Figure \ref{fig:example_1} (g) where, we see that there are no unmatched vertices, so $\mathcal{J}''= \emptyset$.

In step 10, we find the index of one state variable connected to each target-SCC in which that target-SCC has not already been accounted for in $\mathcal{T}$. Hence, $\mathcal{J}'''=~\emptyset$.

Finally, we can combine $\mathcal{J}' \cup \mathcal{J}'' \cup \mathcal{J}''' = \mathcal{J}$ to obtain the minimum set of indices required to place the sensors for achieving a structurally observable system. We find that $\mathcal{J}=\{5,6\}$, and the sensor placement is shown in Figure \ref{fig:example_1} (e) and (f). Since $p=1$, we note that $\mathcal{J}_{x} = \{5,6\}$ and $\mathcal{J}_{d}=\emptyset$. Therefore, we have $\bar C = \mathbb{I}_{n}^{\mathcal{J}_x}$ and $\bar D = \mathbb{I}_{p}^{\mathcal{J}_{d}}$ as the solution to $\mathcal{P}_1$.

We note that there may be circumstances in which it may not be possible to measure the unknown inputs. Therefore, to reduce the number of unknown inputs that need to be measured, it is necessary to avoid the situation where unknown inputs become left unmatched vertices. Hence, we can include a step in the algorithm to place higher weights on the edges in the bipartite graph that connect two vertices without an unknown input as the left vertex as noted in Remark~\ref{avoid_meas_unknown}. 

{\centering
\allowdisplaybreaks
\vspace{2mm} \noindent \colorbox{gray!15}{\parbox{.935\columnwidth}{
\begin{remark}\label{avoid_meas_unknown}
{\small By placing higher weights on the edges in the bipartite graph that do not contain unknown inputs as left vertices, we can ensure that the algorithm avoids measuring unknown inputs where possible.\hfill$\diamond$}
\end{remark}
}
}
}





\section{Special Classes of Systems with Linear-time Computational Complexity}\label{sec:discussion}


In this section, we will outline two classes of switched LTI systems for which a lower computational complexity can be achieved for the minimum sensor placement problem and provide algorithms and examples for these systems. We start by describing two classes of switched LTI systems that allow for a linear-time computational complexity with respect to the edges. The two classes of switched LTI systems are as follows:
\begin{enumerate}
    \item switched LTI systems with unknown inputs that do not have memory and remain constant over time, and the system has nodal dynamics in the states; 
    \item switched LTI systems with nodal dynamics in both inputs and states. 
\end{enumerate} 

\subsection{Switched LTI Systems with memoryless unknown inputs that remain constant over time, and state nodal dynamics}

First, we present an example in which the system has nodal dynamics (i.e., self-loops) in the states, and the unknown inputs do not have memory and remain constant over time. We provide the solution obtained from Algorithm~\ref{alg:main} for this example. Next, we define an algorithm for this class of systems outlined in Algorithm~\ref{alg:class1}. 
\begin{algorithm}[H]
		{\small
		\caption{Dedicated solution for System Class 1}
		\label{alg:class1}
		\begin{algorithmic}[1]
			\STATE{\textbf{input}: A structural switched LTI system with $\mathbb M=\{1,\ldots,m\}$ modes described by $\{\bar A_1,\ldots,\bar A_m,\bar F_1,\hdots,\bar F_m,\bar Q_1,\hdots,\bar Q_m\}$, where $\bar A_k\in\{0,\star\}^{n\times n}$, $\bar F_k\in\{0,\star\}^{n\times p}$, $\bar Q_k\in\{0\}^{p\times p}, \forall k\in\{1,\ldots,m\}$, and $d(t) = c, \forall t$}		\STATE{\textbf{output}: Output $\bar C = \mathbb{\bar I}_n^{\mathcal{J}_{x}}$ and $\bar D= \mathbb{\bar I}_p^{\mathcal{J}_{d}}$, where $\mathcal{J} = \mathcal{J}_{x} \cup \mathcal{J}_{d}$, $\mathcal{J}_{d}=\{i\in\mathcal J:i \leq p\}$, and $\mathcal{J}_{x}=\{i\in\mathcal J:i> p\}$  }
            \STATE{\textbf{set} $\bar A_k'= \text{\small $ 
\begin{bmatrix}
     \mathbf{0}   &  \mathbf{0}   \\[-0.15cm]
     \bar F_k &  \bar A_k 
\end{bmatrix}$} $}

			\STATE{\textbf{find} the $\alpha$ target-SCCs of $\mathcal G\left(\bigvee_{k=1}^m\bar A_k'\right)=(\mathcal X',\mathcal E_{\mathcal X',\mathcal X}')$, denoted by $\{\mathcal S_1,\ldots,\mathcal S_\alpha\}$}
			\STATE{\textbf{Construct} the digraph $\mathcal{G}\equiv\mathcal{G}(\bigvee_{k=1}^{m} \bar A_{k}')$}
			\STATE{\textbf{Extend} $\mathcal{G}$ adding the following nodes: 
			\begin{itemize}
			    \item[$\,\rhd$] a virtual source node $s$;
			    \item[$\,\rhd$] a virtual target node $t$;
			    \item[$\,\rhd$] an ancillary node for each target-SCC denoted by $a_1,\ldots,a_\alpha$. 
			\end{itemize}
			In addition, add the following edges: 
			\begin{itemize}
			    \item[$\,\rhd$] an edge from node $s$ to each unknown input;
			    \item[$\,\rhd$] an edge from each node in the $\mathcal S_i$ to $a_i$, for $i=1,\ldots,\alpha$;
			    \item[$\,\rhd$] an edge from $a_i$ to $t$, for $i=1,\ldots,\alpha$;
			    \item[$\,\rhd$] an edge from each unknown input to $t$. 
			\end{itemize}
			}
			\STATE{\textbf{set} the edge capacities to \\
			{\small
			$\begin{cases}
			1, & \text{if the edge connects the unknown inputs} \\ & \text{to the virtual target node}\\ 
			2, &\text{otherwise}
			\end{cases}
			$
			}
			}
			\STATE{\textbf{find} the vertex-disjoint paths denoted by $\{\mathcal T_1,\ldots,\mathcal T_z\}$} of the digraph constructed in the previous steps starting in the virtual source node $s$ and ending in the virtual target node $t$ using the Ford Fulkerson algorithm to find the maximum flow
			\STATE{\textbf{Set}} $\mathcal J'$ to contain the indices of each node that is connected to the ancillary variables $a_1,\ldots,a_\alpha$ in a disjoint path and the indices of unknown input nodes if they connect to the target node in a disjoint path  
			\STATE{\textbf{Set} $\mathcal J''$ to contain a single index of a state variable from each target-SCC not accounted for in~$\mathcal J'$}
			\STATE{\textbf{Set} $\mathcal J=\mathcal J'\cup\mathcal J''$} 
			\STATE{\textbf{Set} $\mathcal{J}_{d}=\{i\in\mathcal J:i \leq p\}$, and $\mathcal{J}_{x}=\{i\in\mathcal J:i> p\}$ }

		\end{algorithmic} 
		}
\end{algorithm} 
Next, we prove the soundness and derive the computational complexity of Algorithm~\ref{alg:class1}.
\begin{theorem}\label{th:soundness_and_complexity_class1}
Algorithm~\ref{alg:class1} is sound, i.e., it provides a solution to $\mathcal P_1$ for systems in which the unknown inputs do not have memory and remain constant over time, and the system has nodal dynamics in the states. The computational complexity of Algorithm~\ref{alg:class1} is $O((n+p)^2)$, where $n$ is the number of states $x(t)\in\mathbb{R}^{n}$ and $p$ is the number of unknown inputs $d(t)\in\mathbb{R}^{p}$.~\hfill$\circ$
\end{theorem}

We give a brief overview of how the algorithm presented in Algorithm~\ref{alg:class1} differs from Algorithm~\ref{alg:main}. The main difference is in Algorithm~\ref{alg:main} where we must find a MWMM of the bipartite graph $\mathcal{B}([\bar A_1';\ldots;\bar A_m';\bar T])$ whereas in Algorithm~\ref{alg:class1}, we only need to find the vertex-disjoint paths, which is why we can reduce the computational complexity. 

In the next example, we illustrate the results after applying both Algorithm~\ref{alg:main} and Algorithm~\ref{alg:class1}. 

\subsection*{Example 2}

Let us consider the following linear continuous-time system with three modes $\{\bar{A}_{k}, \bar{F}_{k},\bar{Q}_{k}\}_{k=1}^{3}$, where $\bar{A}_{k}\in\{0,1\}^{4\times 4}$, $\bar{F}_{k}\in\{0,1\}^{4\times 1}$, $\bar Q_k\in \{0,1\}$. In particular, 
\begin{equation*}
    A_1 = 
    \left[\begin{smallmatrix}
          1 & 0 & 0 & 0 \\
          0 & 0 & 0 & 0 \\
          1 & 0 & 1 & 0 \\
          0 & 0 & 0 & 0
    \end{smallmatrix}\right]
    , 
    A_2 = 
    \left[\begin{smallmatrix}
          0 & 0 & 0 & 0 \\
          0 & 1 & 0 & 0 \\
          0 & 1 & 1 & 0 \\
          0 & 0 & 0 & 0
          \end{smallmatrix}\right],
    A_3 = 
    \left[\begin{smallmatrix}
          1 & 0 & 0 & 0 \\
          0 & 1 & 0 & 0 \\
          0 & 0 & 1 & 0 \\
          0 & 0 & 1 & 1
          \end{smallmatrix}\right],
\end{equation*}

$[\bar F_{1}]_{1,1} = 1$, $[\bar F_{2}]_{2,1} = 1$, $[\bar F_{3}]_{1,1} = 1$, $[\bar F_{3}]_{2,1} = 1$, and $\bar Q_k=0,\forall k\in\{1,2,3\}$.

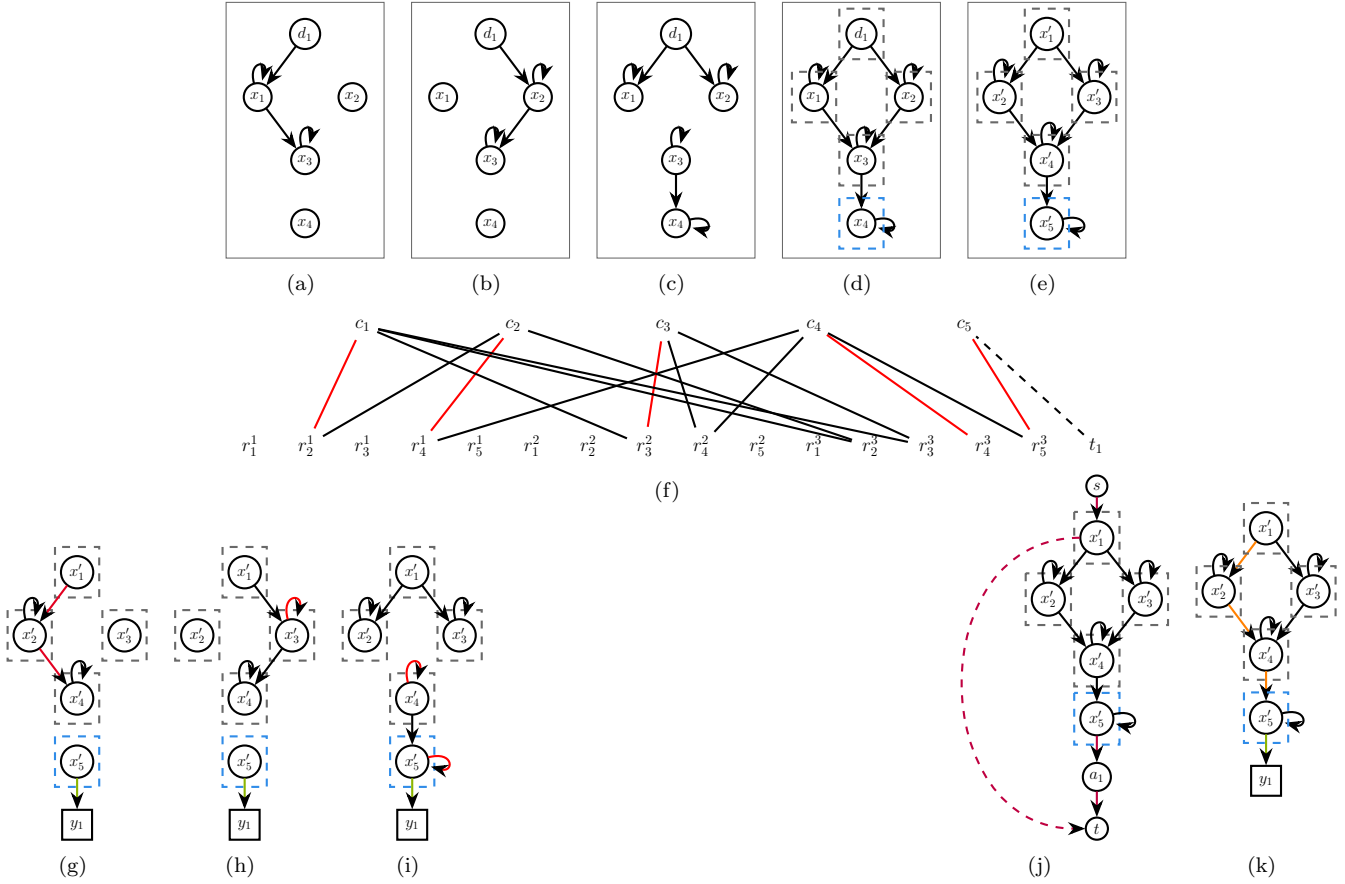
\begin{figure}[H] 
\centering

\subfigure[]{
\begin{tikzpicture}[scale=.42, transform shape,node distance=1.5cm]
\begin{scope}[every node/.style={circle,thick,draw},square/.style={regular polygon,regular polygon sides=4},none/.style={regular polygon,regular polygon sides=1}]
    \node (x1) at (1.5,0) {\Large $d_1$};
    \node (x2) at (0,-2) {\Large $x_1$};
    \node (x3) at (3,-2) {\Large $x_2$};
    \node (x4) at (1.5,-4) {\Large $x_3$};
    \node (x5) at (1.5,-6) {\Large $x_4$};
    
\end{scope}

\begin{scope}[>={Stealth[black]},
              every edge/.style={draw=black, thick},eblue/.style={draw=bleudefrance, thick}]
      
    \path [->] (x1) edge node {} (x2);
    \path [->] (x2) edge node {} (x4);
    \path [->] (x2) edge[loop above] node {} (x2);
    \path [->] (x4) edge[loop above] node {} (x4);

\end{scope}
\draw [draw=dimgray] (-1,1) rectangle (4,-7.1);
\end{tikzpicture}
}
\subfigure[]{
\begin{tikzpicture}[scale=.42, transform shape,node distance=1.5cm]
\begin{scope}[every node/.style={circle,thick,draw},square/.style={regular polygon,regular polygon sides=4},none/.style={regular polygon,regular polygon sides=1}]
    \node (x1) at (1.5,0) {\Large $d_1$};
    \node (x2) at (0,-2) {\Large $x_1$};
    \node (x3) at (3,-2) {\Large $x_2$};
    \node (x4) at (1.5,-4) {\Large $x_3$};
    \node (x5) at (1.5,-6) {\Large $x_4$};
    
\end{scope}

\begin{scope}[>={Stealth[black]},
              every edge/.style={draw=black, thick},eblue/.style={draw=bleudefrance, thick}]
      
    \path [->] (x1) edge node {} (x3);
    \path [->] (x3) edge node {} (x4);
    \path [->] (x3) edge[loop above] node {} (x3);
    \path [->] (x4) edge[loop above] node {} (x4);

\end{scope}
\draw [draw=dimgray] (-1,1) rectangle (4,-7.1);
\end{tikzpicture}
}
\subfigure[]{
\begin{tikzpicture}[scale=.42, transform shape,node distance=1.5cm]
\begin{scope}[every node/.style={circle,thick,draw},square/.style={regular polygon,regular polygon sides=4},none/.style={regular polygon,regular polygon sides=1}]
    \node (x1) at (1.5,0) {\Large $d_1$};
    \node (x2) at (0,-2) {\Large $x_1$};
    \node (x3) at (3,-2) {\Large $x_2$};
    \node (x4) at (1.5,-4) {\Large $x_3$};
    \node (x5) at (1.5,-6) {\Large $x_4$};
    
\end{scope}

\begin{scope}[>={Stealth[black]},
              every edge/.style={draw=black, thick},eblue/.style={draw=bleudefrance, thick}]
      
    \path [->] (x1) edge node {} (x2);
    \path [->] (x1) edge node {} (x3);
    \path [->] (x4) edge node {} (x5);
    \path [->] (x2) edge[loop above] node {} (x2);
    \path [->] (x3) edge[loop above] node {} (x2);
    \path [->] (x4) edge[loop above] node {} (x4);
    \path [->] (x5) edge[loop right] node {} (x5);

\end{scope}
\draw [draw=dimgray] (-1,1) rectangle (4,-7.1);
\end{tikzpicture}
}
\subfigure[]{
\begin{tikzpicture}[scale=.42, transform shape,node distance=1.5cm]
\begin{scope}[every node/.style={circle,thick,draw},square/.style={regular polygon,regular polygon sides=4},none/.style={regular polygon,regular polygon sides=1}]
    \node (x1) at (1.5,0) {\Large $d_1$};
    \node (x2) at (0,-2) {\Large $x_1$};
    \node (x3) at (3,-2) {\Large $x_2$};
    \node (x4) at (1.5,-4) {\Large $x_3$};
    \node (x5) at (1.5,-6) {\Large $x_4$};
    
\end{scope}

\begin{scope}[>={Stealth[black]},
              every edge/.style={draw=black, thick},eblue/.style={draw=bleudefrance, thick}]
      
    \path [->] (x1) edge node {} (x2);
    \path [->] (x1) edge node {} (x3);
    \path [->] (x2) edge node {} (x4);
    \path [->] (x3) edge node {} (x4);
    \path [->] (x4) edge node {} (x5);
    \path [->] (x2) edge[loop above] node {} (x2);
    \path [->] (x3) edge[loop above] node {} (x2);
    \path [->] (x4) edge[loop above] node {} (x4);
    \path [->] (x5) edge[loop right] node {} (x5);

\end{scope}
 \draw [draw=dimgray,dashed,thick] (0.8,0.8) rectangle (2.2,-0.8);
 \draw [draw=dimgray,dashed,thick] (-0.7,-1.2) rectangle (0.7,-2.8);
  \draw [draw=dimgray,dashed,thick] (2.3,-1.2) rectangle (3.7,-2.8);
  \draw [draw=dimgray,dashed,thick] (0.8,-3.2) rectangle (2.2,-4.8);
  \draw [draw=bleudefrance,dashed,thick] (0.8,-5.2) rectangle (2.2,-6.8);
\draw [draw=dimgray] (-1,1) rectangle (4,-7.1);
\end{tikzpicture}
}
\subfigure[]{
\begin{tikzpicture}[scale=.42, transform shape,node distance=1.5cm]
\begin{scope}[every node/.style={circle,thick,draw},square/.style={regular polygon,regular polygon sides=4},none/.style={regular polygon,regular polygon sides=1}]
    \node (x1) at (1.5,0) {\Large $x_1'$};
    \node (x2) at (0,-2) {\Large $x_2'$};
    \node (x3) at (3,-2) {\Large $x_3'$};
    \node (x4) at (1.5,-4) {\Large $x_4'$};
    \node (x5) at (1.5,-6) {\Large $x_5'$};
    
\end{scope}

\begin{scope}[>={Stealth[black]},
              every edge/.style={draw=black, thick},eblue/.style={draw=bleudefrance, thick}]
      
    \path [->] (x1) edge node {} (x2);
    \path [->] (x1) edge node {} (x3);
    \path [->] (x2) edge node {} (x4);
    \path [->] (x3) edge node {} (x4);
    \path [->] (x4) edge node {} (x5);
    \path [->] (x2) edge[loop above] node {} (x2);
    \path [->] (x3) edge[loop above] node {} (x2);
    \path [->] (x4) edge[loop above] node {} (x4);
    \path [->] (x5) edge[loop right] node {} (x5);

\end{scope}
 \draw [draw=dimgray,dashed,thick] (0.8,0.8) rectangle (2.2,-0.8);
 \draw [draw=dimgray,dashed,thick] (-0.7,-1.2) rectangle (0.7,-2.8);
  \draw [draw=dimgray,dashed,thick] (2.3,-1.2) rectangle (3.7,-2.8);
  \draw [draw=dimgray,dashed,thick] (0.8,-3.2) rectangle (2.2,-4.8);
  \draw [draw=bleudefrance,dashed,thick] (0.8,-5.2) rectangle (2.2,-6.8);
\draw [draw=dimgray] (-1,1) rectangle (4,-7.1);
\end{tikzpicture}
}

\vspace{-.2cm}
\subfigure[]{
\begin{tikzpicture}[scale=.5, transform shape,node distance=1.5cm]
\begin{scope}[every node/.style={circle,draw=white},square/.style={regular polygon,regular polygon sides=4},none/.style={regular polygon,regular polygon sides=1}]
    \node (d1_1) at (0,0) {\Large $r_{1}^1$};
    \node (x1_1) at (1.5,0) {\Large $r_{2}^1$};
    \node (x2_1) at (3,0) {\Large $r_{3}^1$};
    \node (x3_1) at (4.5,0) {\Large $r_{4}^1$};
    \node (x4_1) at (6,0) {\Large $r_{5}^1$};
    \node (d1_2) at (7.5,0) {\Large $r_{1}^2$};    
    \node (x1_2) at (9,0) {\Large $r_{2}^2$};
    \node (x2_2) at (10.5,0) {\Large $r_{3}^2$};
    \node (x3_2) at (12,0) {\Large $r_{4}^2$};
    \node (x4_2) at (13.5,0) {\Large $r_{5}^2$};
    \node (d1_3) at (15,0) {\Large $r_{1}^3$};    
    \node (x1_3) at (16.5,0) {\Large $r_{2}^3$};
    \node (x2_3) at (18,0) {\Large $r_{3}^3$};
    \node (x3_3) at (19.5,0) {\Large $r_4^3$};
    \node (x4_3) at (21,0) {\Large $r_5^3$};
    
    \node (t1) at (22.5,0) {\Large $t_1$};
    
    \node (d1) at (3,3.2) {\Large $c_{1}$};
    \node (x1) at (7,3.2) {\Large $c_{2}$};
    \node (x2) at (11,3.2) {\Large $c_{3}$};
    \node (x3) at (15,3.2) {\Large $c_{4}$};
    \node (x4) at (19,3.2) {\Large $c_{5}$};
    
\end{scope}

\begin{scope}[>={Stealth[black]},
              every edge/.style={draw=black, thick},eblue/.style={draw=bleudefrance, thick}]
      
    \path [-] (d1) edge[red] node {} (x1_1);    
    \path [-] (x1) edge node {} (x1_1);
    \path [-] (x1) edge[red] node {} (x3_1);
    \path [-] (x3) edge node {} (x3_1);

    \path [-] (d1) edge node {} (x2_2);  
    \path [-] (x2) edge[red] node {} (x2_2);  
    \path [-] (x2) edge node {} (x3_2);  
    \path [-] (x3) edge node {} (x3_2); 
    
    \path [-] (d1) edge node {} (x1_3);  
    \path [-] (d1) edge node {} (x2_3);  
    \path [-] (x1) edge node {} (x1_3);  
    \path [-] (x2) edge node {} (x2_3);  
    \path [-] (x3) edge[red] node {} (x3_3);  
    \path [-] (x3) edge node {} (x4_3);  
    \path [-] (x4) edge[red] node {} (x4_3); 
    
    \path [-] (t1) edge[dashed] node {} (x4);  

\end{scope}
\end{tikzpicture}
}
\vspace{-0.8cm}

\subfigure[]{
\begin{tikzpicture}[scale=.42, transform shape,node distance=1.5cm]
 \draw [draw=dimgray,dashed,thick] (0.8,0.8) rectangle (2.2,-0.8);
 \draw [draw=dimgray,dashed,thick] (-0.7,-1.2) rectangle (0.7,-2.8);
  \draw [draw=dimgray,dashed,thick] (2.3,-1.2) rectangle (3.7,-2.8);
  \draw [draw=dimgray,dashed,thick] (0.8,-3.2) rectangle (2.2,-4.8);
  \draw [draw=bleudefrance,dashed,thick] (0.8,-5.2) rectangle (2.2,-6.8);
\begin{scope}[every node/.style={circle,thick,draw},square/.style={regular polygon,regular polygon sides=4},none/.style={regular polygon,regular polygon sides=1}]
    \node (x1) at (1.5,0) {\Large $x_1'$};
    \node (x2) at (0,-2) {\Large $x_2'$};
    \node (x3) at (3,-2) {\Large $x_3'$};
    \node (x4) at (1.5,-4) {\Large $x_4'$};
    \node (x5) at (1.5,-6) {\Large $x_5'$};

    \node[square,draw] (y1) at (1.5,-8) {\Large $y_1$};

\end{scope}

\begin{scope}[>={Stealth[black]},
every edge/.style={draw=black, thick},eblue/.style={draw=bleudefrance, thick}]
     \path [->] (x1) edge[cadmiumred] node {} (x2);
    \path [->] (x2) edge[cadmiumred] node {} (x4);
     \path [->] (x5) edge[ao] node {} (y1);
    \path [->] (x2) edge[loop above] node {} (x2);
    \path [->] (x4) edge[loop above] node {} (x4);

\end{scope}
\end{tikzpicture}
}
\subfigure[]{
\begin{tikzpicture}[scale=.42, transform shape,node distance=1.5cm]
 \draw [draw=dimgray,dashed,thick] (0.8,0.8) rectangle (2.2,-0.8);
 \draw [draw=dimgray,dashed,thick] (-0.7,-1.2) rectangle (0.7,-2.8);
  \draw [draw=dimgray,dashed,thick] (2.3,-1.2) rectangle (3.7,-2.8);
  \draw [draw=dimgray,dashed,thick] (0.8,-3.2) rectangle (2.2,-4.8);
  \draw [draw=bleudefrance,dashed,thick] (0.8,-5.2) rectangle (2.2,-6.8);
\begin{scope}[every node/.style={circle,thick,draw},square/.style={regular polygon,regular polygon sides=4},none/.style={regular polygon,regular polygon sides=1}]
    \node (x1) at (1.5,0) {\Large $x_1'$};
    \node (x2) at (0,-2) {\Large $x_2'$};
    \node (x3) at (3,-2) {\Large $x_3'$};
    \node (x4) at (1.5,-4) {\Large $x_4'$};
    \node (x5) at (1.5,-6) {\Large $x_5'$};

    \node[square,draw] (y1) at (1.5,-8) {\Large $y_1$};

\end{scope}

\begin{scope}[>={Stealth[black]},
every edge/.style={draw=black, thick},eblue/.style={draw=bleudefrance, thick}]
    \path [->] (x1) edge node {} (x3);
    \path [->] (x3) edge node {} (x4);
     \path [->] (x5) edge[ao] node {} (y1);
     
    \path [->] (x3) edge[red,loop above] node {} (x3);
    \path [->] (x4) edge[loop above] node {} (x4);

\end{scope}
\end{tikzpicture}
}
\subfigure[]{
\begin{tikzpicture}[scale=.42, transform shape,node distance=1.5cm]
 \draw [draw=dimgray,dashed,thick] (0.8,0.8) rectangle (2.2,-0.8);
 \draw [draw=dimgray,dashed,thick] (-0.7,-1.2) rectangle (0.7,-2.8);
  \draw [draw=dimgray,dashed,thick] (2.3,-1.2) rectangle (3.7,-2.8);
  \draw [draw=dimgray,dashed,thick] (0.8,-3.2) rectangle (2.2,-4.8);
  \draw [draw=bleudefrance,dashed,thick] (0.8,-5.2) rectangle (2.2,-6.8);
\begin{scope}[every node/.style={circle,thick,draw},square/.style={regular polygon,regular polygon sides=4},none/.style={regular polygon,regular polygon sides=1}]
    \node (x1) at (1.5,0) {\Large $x_1'$};
    \node (x2) at (0,-2) {\Large $x_2'$};
    \node (x3) at (3,-2) {\Large $x_3'$};
    \node (x4) at (1.5,-4) {\Large $x_4'$};
    \node (x5) at (1.5,-6) {\Large $x_5'$};

    \node[square,draw] (y1) at (1.5,-8) {\Large $y_1$};

\end{scope}

\begin{scope}[>={Stealth[black]},
every edge/.style={draw=black, thick},eblue/.style={draw=bleudefrance, thick}]
    \path [->] (x1) edge node {} (x3);
    \path [->] (x1) edge node {} (x2);
    \path [->] (x4) edge node {} (x5);
     \path [->] (x5) edge[ao] node {} (y1);
     
    \path [->] (x3) edge[loop above] node {} (x3);
    \path [->] (x2) edge[loop above] node {} (x2);
    \path [->] (x4) edge[red,loop above] node {} (x4);
    \path [->] (x5) edge[red,loop right] node {} (x5);

\end{scope}
\end{tikzpicture}
}
\hfill
\subfigure[]{
\begin{tikzpicture}[scale=.43, transform shape,node distance=1.5cm]
 \draw [draw=dimgray,dashed,thick] (0.8,0.8) rectangle (2.2,-0.8);
 \draw [draw=dimgray,dashed,thick] (-0.7,-1.1) rectangle (0.7,-2.7);
  \draw [draw=dimgray,dashed,thick] (2.3,-1.1) rectangle (3.7,-2.7);
  \draw [draw=dimgray,dashed,thick] (0.8,-3) rectangle (2.2,-4.6);
  \draw [draw=bleudefrance,dashed,thick] (0.8,-4.8) rectangle (2.2,-6.4);
\begin{scope}[every node/.style={circle,thick,draw},square/.style={regular polygon,regular polygon sides=4},none/.style={regular polygon,regular polygon sides=1}]
    \node (s) at (1.5, 1.6) {\Large $s$};
    \node (x1) at (1.5,0) {\Large $x_1'$};
    \node (x2) at (0,-1.9) {\Large $x_2'$};
    \node (x3) at (3,-1.9) {\Large $x_3'$};
    \node (x4) at (1.5,-3.8) {\Large $x_4'$};
    \node (x5) at (1.5,-5.6) {\Large $x_5'$};
    \node (a1) at (1.5, -7.4) {\Large $a_1$};
    \node (t) at (1.5, -9) {\Large $t$};

\end{scope}

\begin{scope}[>={Stealth[black]},
every edge/.style={draw=black, thick},eblue/.style={draw=bleudefrance, thick}]
    \path [->] (s) edge[purple] node {} (x1);
    \path [->] (x1) edge node {} (x3);
    \path [->] (x1) edge node {} (x2);
    \path [->] (x2) edge node {} (x4);
    \path [->] (x3) edge node {} (x4);
    \path [->] (x4) edge node {} (x5);
    \path [->] (x5) edge[purple] node {} (a1);
    \path [->] (a1) edge[purple] node {} (t);
    \path[->] (x1) edge[purple, dashed, in=180,out=-180,looseness=1.4] node {} (t);

    \path [->] (x3) edge[loop above] node {} (x3);
    \path [->] (x2) edge[loop above] node {} (x2);
    \path [->] (x4) edge[loop above] node {} (x4);
    \path [->] (x5) edge[loop right] node {} (x5);

\end{scope}
\end{tikzpicture}
}
\subfigure[]{
\begin{tikzpicture}[scale=.42, transform shape,node distance=1.5cm]
\begin{scope}[every node/.style={circle,thick,draw},square/.style={regular polygon,regular polygon sides=4},none/.style={regular polygon,regular polygon sides=1}]
    \node (x1) at (1.5,0) {\Large $x_1'$};
    \node (x2) at (0,-2) {\Large $x_2'$};
    \node (x3) at (3,-2) {\Large $x_3'$};
    \node (x4) at (1.5,-4) {\Large $x_4'$};
    \node (x5) at (1.5,-6) {\Large $x_5'$};
    \node[square,draw] (y1) at (1.5,-8) {\Large $y_1$};
    \node[draw=white] (s) at (1.5, 1.6) { };
    \node[draw=white] (t) at (1.5, -9.7) { };
\end{scope}

\begin{scope}[>={Stealth[black]},
              every edge/.style={draw=black, thick},eblue/.style={draw=bleudefrance, thick}]
      
    \path [->] (x1) edge[orange(webcolor)] node {} (x2);
    \path [->] (x1) edge node {} (x3);
    \path [->] (x2) edge[orange(webcolor)] node {} (x4);
    \path [->] (x3) edge node {} (x4);
    \path [->] (x4) edge[orange(webcolor)] node {} (x5);
    \path [->] (x2) edge[loop above] node {} (x2);
    \path [->] (x3) edge[loop above] node {} (x3);
    \path [->] (x4) edge[loop above] node {} (x4);
    \path [->] (x5) edge[loop right] node {} (x5);
     \path [->] (x5) edge[ao] node {} (y1);

\end{scope}
 \draw [draw=dimgray,dashed,thick] (0.8,0.8) rectangle (2.2,-0.8);
 \draw [draw=dimgray,dashed,thick] (-0.7,-1.2) rectangle (0.7,-2.8);
  \draw [draw=dimgray,dashed,thick] (2.3,-1.2) rectangle (3.7,-2.8);
  \draw [draw=dimgray,dashed,thick] (0.8,-3.2) rectangle (2.2,-4.8);
  \draw [draw=bleudefrance,dashed,thick] (0.8,-5.2) rectangle (2.2,-6.8);
\end{tikzpicture}
}
\vspace{-0.3cm}
\caption{ (a) shows $A_{1}'$, (b) shows $A_{2}'$ (c) shows $A_{3}'$ and (d) shows the union of the three modes of the continuous-time system with unknown inputs (e) shows the augmented continuous-time system. (f) shows the bipartite graph $\mathcal{B}([\bar{A}_{1}';\bar{A}_{2}';\bar{A}_{3}';\bar T])$ with unitary weight on the dotted edge and zero weight on the solid edges. The collection of red edges is the minimum weight maximum matching. (g)-(i) show the SCCs in dotted black rectangles, the target-SCC in a dotted blue rectangle, and the minimal output sensors and their placement for all three modes after applying Algorithm~\ref{alg:main} (j) shows the digraph with the extra nodes and edges (shown in purple) added in steps~6 and ~7. The dashed purple line signifies that the edge has a weight of 2. (k) shows the vertex disjoint path in orange and the solution for the minimal output sensors placement after applying Algorithm~\ref{alg:class1}.}
\label{fig:example_2}
\end{figure}

The individual modes of the system are shown in Figure \ref{fig:example_2} (a)-(c). We apply Algorithm \ref{alg:main} to this system to find the minimum set of dedicated sensors to achieve structural observability. We start by finding the union of the modes. $\bigvee_{k=1}^{m}\bar{A}_{k}'$ is given as follows
\begin{equation*}
    \bar{A}_{1}'\vee\bar{A}_{2}'\vee\bar{A}_{3}'= 
    \text{\small $
    \left[\begin{array}{@{}c|cccc@{}}
    0 & 0 & 0 & 0 & 0 \\ \hline
    1 & 1 & 0 & 0 & 0 \\
    1 & 0 & 1 & 0 & 0 \\
    0 & 1 & 1 & 1 & 0 \\
    0 & 0 & 0 & 1 & 1 
  \end{array}\right]
  $}
\end{equation*}
and is shown in Figure \ref{fig:example_2} (d) and (e), where (d) shows the union of the modes before the system has been augmented and (e) shows the union of the modes after the system has been augmented and relabeled with state $x'=[d^{\intercal}$  $x^{\intercal}]^{\intercal}$. With the system properly combined and augmented, we continue by finding the target-SCCs of $\mathcal{G}(\bigvee_{k=1}^{m}\bar{A}_{k}')$. We find that there are 5 SCCs, which are outlined in dashed rectangle boxes in Figure \ref{fig:example_2} (d)-(e). We also find that there is 1 target-SCC, which is outlined in a dashed blue box in Figure \ref{fig:example_2} (d)-(e), so $\alpha=1$.

We construct the bipartite graph $\mathcal{B}([\bar{A}_{1}';\bar{A}_{2}';\bar{A}_{3}';\bar T])$. First, we note that since there is only one target-SCC composed of one state ($x_5'$), then $
    \bar T = \begin{bmatrix}
    0 & 0 & 0 & 0 & 1
    \end{bmatrix}.$ 
In the bipartite graph depicted in Figure \ref{fig:example_2} (f), there is one edge that connects to the target-SCC and thus has unitary weight (shown as a dotted line) while the rest of the edges have zero weight (shown as solid lines). In step 7, we find the MWMM $\mathcal{M}'$. There are different possible MWMM, one of which is shown by the collection of red edges in Figures \ref{fig:example_2} (f). From the MWMM, we see that there aren't any edges in the MWMM that are connected to the target-SCC, so $\mathcal{J}' = \emptyset$. In step 9, for the MWMM in Figure \ref{fig:example_2} (f), we find $\mathcal{J}'' = \emptyset$, which is the set of indices associated with the column vertices of $\bar A_k', \forall k$ that are not in the MWMM (i.e., all of the indices $j=\{1,\dots,n+p\}$ such that $c_{j}$ are not connected to a red edge in Figure \ref{fig:example_2} (f)). 

In step 10, we find the index of one state variable connected to each target-SCC for each not already accounted target-SCC. Hence, $\mathcal{J}'''=\{5\}$.

Finally, we can combine $\mathcal{J}' \cup \mathcal{J}'' \cup \mathcal{J}''' = \mathcal{J}=\{5\}$ to obtain the minimum set of indices required to place the sensors for achieving a structurally observable system.  Since $p=1$, we find that $\mathcal{J}_{x} = \{5\}$ and $\mathcal{J}_{d}=\emptyset$.
Therefore, we have $\bar C = \mathbb{I}_{n}^{\mathcal{J}_x}$ and $\bar D = \mathbb{I}_{p}^{\mathcal{J}_{d}}$.

Next, we apply Algorithm~\ref{alg:class1} to this system. We again start by finding the target-SCCs of $\mathcal G(\bigvee_{k=1}^{m}\bar A_{k}')$, which are outlined in blue dashed boxes in Figure \ref{fig:example_2} (d)-(e). Then, we construct the system digraph $\mathcal{G}(\bigvee_{k=1}^{m} \bar A_{k}')$ along with the additional virtual nodes, ancillary nodes, edges, and appropriate edge capacities described in step~7, which is shown in Figure \ref{fig:example_2} (j). Next, we find the vertex-disjoint paths of the digraph pictured in Figure \ref{fig:example_2} (j), which are shown as orange paths in Figures \ref{fig:example_2} (k). We find one vertex disjoint path with ending in node 5, so $\mathcal{J}' = \{5\}$ (see Figure \ref{fig:example_2} (k)). Since there is only one target-SCC, which has already been accounted for in $\mathcal{J}'$, then we find that $\mathcal J'' = \emptyset$. Combining both sets, we find the same solution as calculated in Algorithm~\ref{alg:main}, namely $\mathcal J = \mathcal J' \cup \mathcal J'' = \{5\}$. Finally, this results in the same $\mathcal J_d$ and $\mathcal J_x$ sets as well.     

{\centering
\noindent \colorbox{gray!15}{\parbox{.935\columnwidth}{
\begin{remark}\label{linear_time}
{\small In~\cite{kawarabayashi2012disjoint}, the authors present a quadratic (in the number of vertices) time-complexity algorithm to compute a set of independent paths in a digraph between pairs of given vertices. 
The computational complexity for finding the induced disjoint paths problem can be reduced to linear-time (in the number of vertices) if the graph is planar as seen in ~\cite{kawarabayashi2012linear}. A digraph is \emph{planar} if it can be drawn in the plane without any edge intersections. As such, finding the minimum sensors in Examples 2 and 3 can be reduced to a linear-time  computational complexity with respect to the number of vertices plus edges.\hfill$\diamond$} 
\end{remark}}}
}



\subsection{Systems with Nodal Dynamics}

For digraphs that are spanned by cycles, finding the minimum sensor placement only requires finding the target-SCCs \cite{pequito2015framework}. Next, we present an example where the digraph is not only spanned by cycles but has nodal dynamics meaning every state has a self-loop. We will show through this example that the conditions for finding the minimum number of dedicated sensors to ensure structural state and input observability for digraphs spanned by cycles boils down to finding the target-SCCs. We present Algorithm~\ref{alg:class2}, which finds the dedicated solution for the System Class 2 (i.e., nodal dynamics). 

\begin{algorithm}[H]
		{\small
		\caption{Dedicated solution for System Class 2}
		\label{alg:class2}
		\begin{algorithmic}[1]
			\STATE{\textbf{input}: A structural switched LTI system with $\mathbb M=\{1,\ldots,m\}$ modes described by $\{\bar A_1,\ldots,\bar A_m,\bar F_1,\hdots,\bar F_m,\bar Q_1,\hdots,\bar Q_m\}$, where $\bar A_k\in\{0,\star\}^{n\times n}$, $\bar F_k\in\{0,\star\}^{n\times p}$, $\bar Q_k\in\{0,\star\}^{p\times p},$, and the diagonal entries of $A_k$ and $Q_k$ are nonzero for all $k\in\{1,\ldots,m\}$}		\STATE{\textbf{output}: Output $\bar C = \mathbb{\bar I}_n^{\mathcal{J}_{x}}$ and $\bar D= \mathbb{\bar I}_p^{\mathcal{J}_{d}}$, where $\mathcal{J} = \mathcal{J}_{x} \cup \mathcal{J}_{d}$, $\mathcal{J}_{d}=\{i\in\mathcal J:i \leq p\}$, and $\mathcal{J}_{x}=\{i\in\mathcal J:i> p\}$}
            \STATE{\textbf{set} $\bar A_k'=
            \left[\begin{smallmatrix}
     \bar Q_k   &  \mathbf{0}   \\
     \bar F_k &  \bar A_k 
\end{smallmatrix}\right]
$
            }
			\STATE{\textbf{find} the $\alpha$ target-SCCs of $\mathcal G\left(\bigvee_{k=1}^m\bar A_k'\right)=(\mathcal X',\mathcal E_{\mathcal X',\mathcal X'})$, denoted by $\{\mathcal S_1,\ldots,\mathcal S_\alpha\}$}
			\STATE{\textbf{Set} $\mathcal J$ to contain one and only one index of a state variable from each target-SCC in $\{\mathcal S_s\,:\,s\in\{1,\ldots,\alpha\}\}$} 
			\STATE{\textbf{Set} $\mathcal{J}_{d}=\{i\in\mathcal J:i \leq p\}$, and $\mathcal{J}_{x}=\{i\in\mathcal J:i> p\}$ }
		\end{algorithmic} 
		}
\end{algorithm}

\begin{theorem}\label{th:soundness_and_complexity_class2}
Algorithm~\ref{alg:class2} is sound, i.e., it provides a solution to $\mathcal P_1$ for systems with nodal dynamics. The computational complexity of Algorithm~\ref{alg:class2} is $O((n+p)^2)$, where $n$ is the number of states $x(t)\in\mathbb{R}^{n}$, and $p$ is the number of unknown inputs.~\hfill$\circ$
\end{theorem}

We note in Remark~\ref{distributed} that the solution for this example can be found in a distributed fashion.

{\centering
\allowdisplaybreaks
\noindent \colorbox{gray!15}{\parbox{.935\columnwidth}{
\begin{remark}\label{distributed}
{\small We remark that a distributed solution can be used to find the minimum solution for nodal systems (i.e., a system with self-loops)  \cite{pequito2015distributed}.\hfill$\diamond$
}
\end{remark}
}
}
}

We give a brief overview of how the algorithm presented in Algorithm~\ref{alg:class2} differs from Algorithm~\ref{alg:main}. The main difference is in Algorithm~\ref{alg:class2} we only need to find the target-SCCs, which is why we can reduce the computational complexity. 

In the next example, we illustrate the results after applying both Algorithm~\ref{alg:main} and Algorithm~\ref{alg:class2} to a system with nodal dynamics. 

Let us consider the following linear continuous-time system with two modes $\{\bar{A}_{k}, \bar{F}_{k},\bar{Q}_{k}\}_{k=1}^{2}$, where $\bar{A}_{k}\in\{0,1\}^{2\times 2}$, $\bar{F}_{k}\in\{0,1\}^{2\times 1}$, $\bar Q_k\in \{0,1\}$. In particular, $[\bar A_{1}]_{1,1} = 1$, $[\bar A_{2}]_{2,2} = 1$, $[\bar F_{1}]_{1,1} = 1$, $[\bar F_{2}]_{2,1} = 1$, and $\bar Q_k=1,\forall i\in\{1,2\}$.

\begin{figure}[H] 
\centering

\subfigure[]{
\begin{tikzpicture}[scale=.47, transform shape,node distance=1.5cm]
\begin{scope}[every node/.style={circle,thick,draw},square/.style={regular polygon,regular polygon sides=4},none/.style={regular polygon,regular polygon sides=1}]
    \node (x1) at (1.5,0) {\Large $d_1$};
    \node (x2) at (0,-2) {\Large $x_1$};
    \node (x3) at (3,-2) {\Large $x_2$};
    
\end{scope}

\begin{scope}[>={Stealth[black]},
              every edge/.style={draw=black, thick},eblue/.style={draw=bleudefrance, thick}]
      
    \path [->] (x1) edge node {} (x2);
    \path [->] (x1) edge[loop right] node {} (x1);
    \path [->] (x2) edge[loop above] node {} (x2);

\end{scope}
 \draw [draw=dimgray] (-0.8,0.8) rectangle (3.8,-2.8);
\end{tikzpicture}
}
\subfigure[]{
\begin{tikzpicture}[scale=.47, transform shape,node distance=1.5cm]
\begin{scope}[every node/.style={circle,thick,draw},square/.style={regular polygon,regular polygon sides=4},none/.style={regular polygon,regular polygon sides=1}]
    \node (x1) at (1.5,0) {\Large $d_1$};
    \node (x2) at (0,-2) {\Large $x_1$};
    \node (x3) at (3,-2) {\Large $x_2$};
    
\end{scope}

\begin{scope}[>={Stealth[black]},
              every edge/.style={draw=black, thick},eblue/.style={draw=bleudefrance, thick}]
      
    \path [->] (x1) edge node {} (x3);
    \path [->] (x3) edge[loop above] node {} (x3);
    \path [->] (x1) edge[loop right] node {} (x1);

\end{scope}
 \draw [draw=dimgray] (-0.8,0.8) rectangle (3.8,-2.8);
\end{tikzpicture}
}
\subfigure[]{
\begin{tikzpicture}[scale=.47, transform shape,node distance=1.5cm]
\begin{scope}[every node/.style={circle,thick,draw},square/.style={regular polygon,regular polygon sides=4},none/.style={regular polygon,regular polygon sides=1}]
    \node (x1) at (1.5,0) {\Large $d_1$};
    \node (x2) at (0,-2) {\Large $x_1$};
    \node (x3) at (3,-2) {\Large $x_2$};   
\end{scope}

\begin{scope}[>={Stealth[black]},
              every edge/.style={draw=black, thick},eblue/.style={draw=bleudefrance, thick}]
      
    \path [->] (x1) edge node {} (x3);
    \path [->] (x3) edge[loop above] node {} (x3);
    \path [->] (x1) edge[loop right] node {} (x1);
    \path [->] (x1) edge node {} (x2);
    \path [->] (x2) edge[loop above] node {} (x2);

\end{scope}
 \draw [draw=dimgray] (-0.8,0.8) rectangle (3.8,-2.8);
\end{tikzpicture}
}

\vspace{-0.2cm}
\subfigure[]{
\begin{tikzpicture}[scale=.47, transform shape,node distance=1.5cm]
\begin{scope}[every node/.style={circle,thick,draw},square/.style={regular polygon,regular polygon sides=4},none/.style={regular polygon,regular polygon sides=1}]
    \node (x1) at (1.5,0) {\Large $x_1'$};
    \node (x2) at (0,-2) {\Large $x_2'$};
    \node (x3) at (3,-2) {\Large $x_3'$};   
\end{scope}

\begin{scope}[>={Stealth[black]},
              every edge/.style={draw=black, thick},eblue/.style={draw=bleudefrance, thick}]
      
    \path [->] (x1) edge node {} (x3);
    \path [->] (x3) edge[loop above] node {} (x3);
    \path [->] (x1) edge[loop right] node {} (x1);
    \path [->] (x1) edge node {} (x2);
    \path [->] (x2) edge[loop above] node {} (x2);

\end{scope}
 \draw [draw=dimgray] (-0.8,0.8) rectangle (3.8,-2.8);
\end{tikzpicture}
}
\subfigure[]{
\begin{tikzpicture}[scale=.43, transform shape,node distance=1.5cm]
\draw [draw=dimgray,dashed,thick] (0.8,0.7) rectangle (2.2,-0.7);
  \draw [draw=bleudefrance,dashed,thick] (-0.7,-1.3) rectangle (0.7,-2.7);
  \draw [draw=bleudefrance,dashed,thick] (2.3,-1.3) rectangle (3.7,-2.7);

\begin{scope}[every node/.style={circle,thick,draw},square/.style={regular polygon,regular polygon sides=4},none/.style={regular polygon,regular polygon sides=1}]
    \node (x1) at (1.5,0) {\Large $x_1'$};
    \node (x2) at (0,-2) {\Large $x_2'$};
    \node (x3) at (3,-2) {\Large $x_3'$};   
    \node[square,draw] (y1) at (0,-3.9) {\Large $y_1$};
    \node[square,draw] (y2) at (3,-3.9) {\Large $y_2$};
\end{scope}

\begin{scope}[>={Stealth[black]},
              every edge/.style={draw=black, thick},eblue/.style={draw=bleudefrance, thick}]
      
    \path [->] (x1) edge node {} (x3);
    \path [->] (x3) edge[loop above,cadmiumred] node {} (x3);
    \path [->] (x1) edge[loop right,cadmiumred] node {} (x1);
    \path [->] (x1) edge node {} (x2);
    \path [->] (x2) edge[loop above,cadmiumred] node {} (x2);
    \path [->] (x2) edge[ao] node {} (y1);
     \path [->] (x3) edge[ao] node {} (y2);

\end{scope}

\end{tikzpicture}
}

\vspace{-.3cm}
\subfigure[]{
\begin{tikzpicture}[scale=.5, transform shape,node distance=1.5cm]
\begin{scope}[every node/.style={circle,draw=white},square/.style={regular polygon,regular polygon sides=4},none/.style={regular polygon,regular polygon sides=1}]
    \node (d1_1) at (0,0) {\Large $r_{1}^1$};
    \node (x1_1) at (1.2,0) {\Large $r_{2}^1$};
    \node (x2_1) at (2.4,0) {\Large $r_{3}^1$};
    \node (d1_2) at (3.6,0) {\Large $r_{1}^2$};    
    \node (x1_2) at (4.8,0) {\Large $r_{2}^2$};
    \node (x2_2) at (6,0) {\Large $r_{3}^2$};

    \node (t1) at (7.2,0) {\Large $t_1$};
    \node (t2) at (8.4,0) {\Large $t_2$};
    
    \node (d1) at (2,2) {\Large $c_{1}$};
    \node (x1) at (4.1,2) {\Large $c_{2}$};
    \node (x2) at (6.2,2) {\Large $c_{3}$};
    
\end{scope}

\begin{scope}[>={Stealth[black]},
              every edge/.style={draw=black, thick},eblue/.style={draw=bleudefrance, thick}]
      
    \path [-] (d1) edge[red] node {} (d1_1);    
    \path [-] (d1) edge node {} (x1_1);    
    \path [-] (x1) edge[red] node {} (x1_1);  
    
    \path [-] (d1) edge node {} (d1_2);
    \path [-] (d1) edge node {} (x2_2);
    \path [-] (x2) edge[red] node {} (x2_2);
    
    \path [-] (t1) edge[dashed] node {} (x1);
    \path [-] (t2) edge[dashed] node {} (x2);

\end{scope}
\end{tikzpicture}
}

\vspace{-0.4cm}
\caption{(a)  shows $A_{1}'$ (b) shows $A_{2}'$ (c) shows the union of the two modes of the continuous-time system with unknown inputs, and (d) shows the augmented continuous-time system. (e) shows the SCCs in dotted black rectangles, the target-SCCs in a dotted blue rectangle, and the minimal output sensors and their placement. (f) shows the bipartite graph $\mathcal{B}([\bar{A}_{1}';\bar{A}_{2}';\bar T])$ with unitary weight on the dotted edge and zero weight on the solid edges. The collection of red edges is the minimum weight maximum matching. }
\label{fig:example_loops}
\end{figure}
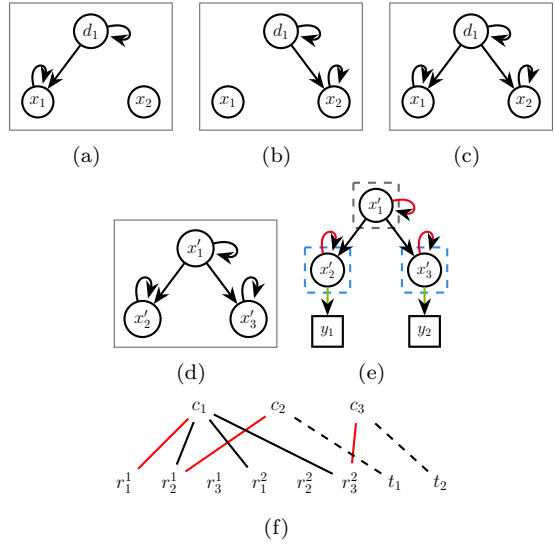

The individual modes of the system are shown in Figure \ref{fig:example_loops} (a)-(b). We apply Algorithm \ref{alg:main} to this system to find the minimum set of dedicated sensors to achieve structural observability. We start by finding the union of the modes. 
$\displaystyle\bigvee_{k=1}^{m}\bar{A}_{k}'$ is given as 
$
    \bar{A}_{1}'\vee\bar{A}_{2}'= 
    \text{\small $
    \left[\begin{array}{@{}c|cc@{}}
    1 & 0 & 0 \\
    \hline
    1 & 1 & 0 \\
    1 & 0 & 1 
  \end{array}
  \right]
  $}
$
and is shown in Figure \ref{fig:example_loops} (c) and (d), where (c) shows the union of the modes before the system has been augmented and (d) shows the union of the modes after the system has been augmented and relabeled with state $x'={[d^{\intercal}\,\,  x^{\intercal}]}^{\intercal}$. With the system properly combined and augmented, we continue by finding the target-SCCs of $\mathcal{G}(\bigvee{k=1}^{m}\bar{A}_{k}')$. We find that there are 3 SCCs, which are outlined in dashed rectangle boxes in Figure \ref{fig:example_loops} (e). We also find that there are 2 target-SCCs, which are outlined in a dashed blue box in Figure \ref{fig:example_loops} (e), so $\alpha=2$.

We construct the bipartite graph $\mathcal{B}([\bar{A}_{1}';\bar{A}_{2}';\bar T])$. Since there are two target-SCCs each composed of one state ($x_{2}'$, $x_3'$), then $\bar T$ is as $
    \bar T = 
    \left[\begin{smallmatrix}
    0 & 0 & 1\\
    0 & 1 & 0
    \end{smallmatrix}\right]
    .$ 
In the bipartite graph depicted in Figure \ref{fig:example_loops} (f), there are two edges that each connect to one of the two target-SCCs and thus have unitary weight (shown as a dotted line) while the rest of the edges have zero weight (shown as solid lines). In step 7, we find the MWMM $\mathcal{M}'$, which is shown by the collection of red edges in Figure \ref{fig:example_loops} (f). 
From the MWMM, we see that there aren't any edges in the MWMM that are connected to a target-SCC, so $\mathcal{J}' = \emptyset$. In step 9, we find $\mathcal{J}''$, which is the set of indices associated with the column vertices $\bar A'$ that are not in the MWMM (i.e., all of the indices $j=\{1,\dots,n+p\}$ such that $c_{j}$ are not connected to a red edge in Figure \ref{fig:example_loops} (f)).
We find that $\mathcal{J}''=\emptyset$.
In step 10, we find the index of one state variable connected to each target-SCC for each not already accounted target-SCC. Hence, $\mathcal{J}'''=\{2,3\}$.

Finally, we can combine $\mathcal{J}' \cup \mathcal{J}'' \cup \mathcal{J}''' = \mathcal{J}=\{2,3\}$ to obtain the minimum set of indices required to place the sensors for achieving a structurally observable system. The sensor placement is shown in Figure \ref{fig:example_loops} (e). 
Since $p=1$, we note that $\mathcal{J}_{x} = \{2,3\}$ and $\mathcal{J}_{d}=\emptyset$. Therefore, we have $\bar C = \mathbb{I}_{n}^{\mathcal{J}_x}$ and $\bar D = \mathbb{I}_{p}^{\mathcal{J}_{d}}$.

Next, we apply Algorithm~\ref{alg:class2} to this system. After augmenting the system, we again start by finding the target-SCCs of $\mathcal G(\bigvee_{k=1}^{m} \bar A_{k}')$, which are outlined in the dashed blue boxes in Figure \ref{fig:example_loops} (f). Next, we find a single state in each of the two target-SCCs and add their indices to $\mathcal{J}$. Hence, $\mathcal J=\{2,3\}$, which we note is the same result achieved by Algorithm~\ref{alg:main}. 

\section{Real-World Example}\label{sec:real_world}
In this section, we find the minimum sensor placement for a real-world example from power systems by considering the IEEE 5-bus system \cite{ramos2013model}, which has three generators and two loads. Through linearization, we can model this system as a continuous-time switched LTI system with unknown inputs by considering two modes. One mode is the working system, and the second mode contains a fault that disconnects generator 1 to load 1, which corresponds to the connection between $x_{14}$ and $x_{10}$ being eliminated. The unknown inputs $d_1$ and $d_2$ capture the unknown amount of load consumed by loads 1 and 2, respectively. Table \ref{tab:ieee_5_bus} describes the states and unknown inputs of the network. The shaded rows in the table correspond to the unknown inputs. The variables/nodes that are not listed in the table but appear in the system digraph correspond to the internal variables that connect the different bus, generators, and loads. The blue nodes correspond to load 1. The orange nodes correspond to load 2. The green nodes correspond to generator 1. The red nodes correspond to generator 2. The gray nodes correspond to generator 3.

\begin{table}[H]
\centering
{\footnotesize
    \begin{tabular}{|c|c|}
        \hline \textbf{Description} & \textbf{Node} \\\hline
         frequency of $G_1$ & $x_1$ \\ \hline
         turbine output mechanical power of $G_1$ & $x_2$ \\\hline
         steam valve opening position of $G_1$ & $x_3$ \\ \hline
         frequency of $G_2$ & $x_4$ \\ \hline
         turbine output mechanical power of $G_2$ & $x_5$ \\\hline
         steam valve opening position of $G_2$ & $x_6$ \\ \hline
         frequency of $G_3$ & $x_7$ \\ \hline
         turbine output mechanical power of $G_3$ & $x_8$ \\\hline
         steam valve opening position of $G_3$ & $x_9$  \\ \hline
         \rowcolor{blue!50}
         unknown uncertainty $L_1$ & $d_1$ \\\hline
         load consumed by $L_1$ & $x_{10}$ \\\hline
         \rowcolor{orange!50}
         unknown uncertainty of $L_2$ & $d_2$ \\\hline
         load consumed by $L_2$ & $x_{12}$ \\\hline
    \end{tabular}
    }
    \caption{States and Unknown Inputs for IEEE 5-bus system}
    \label{tab:ieee_5_bus}
\end{table}

\begin{figure}[H]
\centering
\includegraphics[width=.5\linewidth]{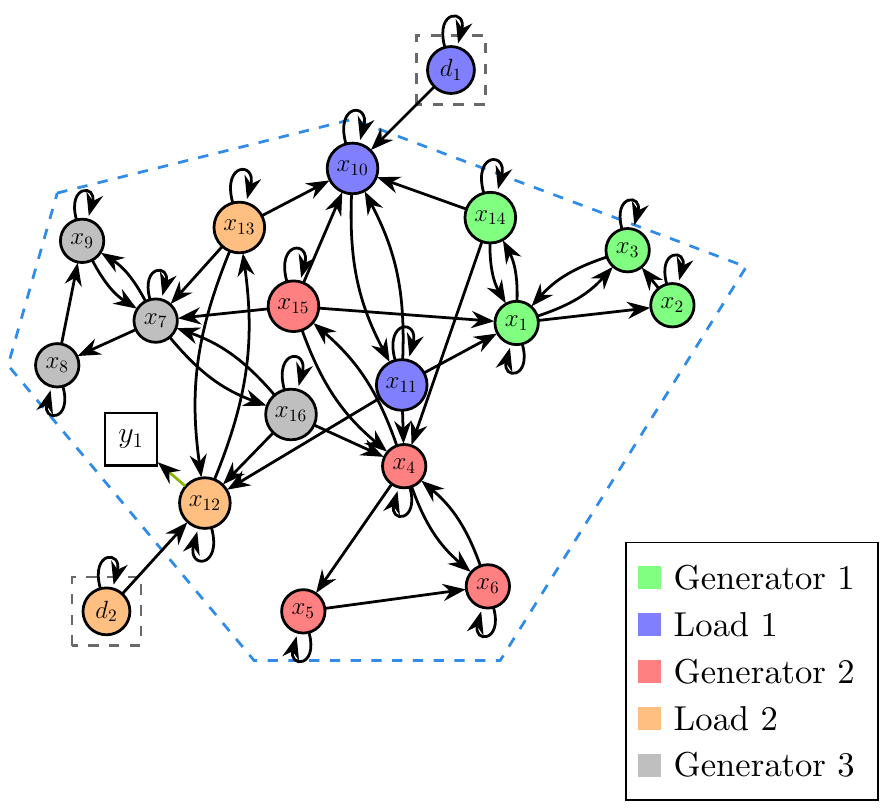}
\caption{This figure shows the union of the two modes of the continuous-time system with unknowns for the IEEE 5-bus system. The SCCs are outlined by two dotted black rectangles, and the target-SCCs is outlined by a dotted blue polygon. The minimum output sensor and its placement is shown by $y_1$.}
\label{fig:ieee5bus}
\end{figure}

The union of the two modes are shown in Figure~\ref{fig:ieee5bus}. Since the system possesses nodal dynamics on all both the inputs and states, we apply Algorithm~\ref{alg:class2} to this system to find the minimum set of dedicated sensors to achieve structural observability. We start by finding the union of the modes, which is shown in Figure~\ref{fig:ieee5bus}. Next, we augmented the system and relabeled it with state $x'={[d^{\intercal}\,\,  x^{\intercal}]}^{\intercal}$. With the system properly combined and augmented, we continue by finding the target-SCCs of $\mathcal{G}(\bigvee{k=1}^{m}\bar{A}_{k}')$. We find that there are 3 SCCs, which are outlined in dashed polygons in Figure~\ref{fig:ieee5bus}. We also find that there is 1 target-SCCs, which is outlined in a dashed blue polygon in Figure~\ref{fig:ieee5bus}, so $\alpha=1$. Next, we find a single state in the target-SCCs and add its indices to $\mathcal{J}$. In this real-world example, it makes most sense to measure the load consumed from either load. Hence, $\mathcal J=\{12\}$ or $\mathcal J=\{10\}$. 

\section{Conclusions}\label{sec:conclusions}
In this paper, we investigated the structural state and input observability of continuous-time switched LTI systems under unknown inputs when unknown inputs can be modeled as LTI systems. 
To this end, we derived necessary and sufficient conditions for the structural state and input observability of continuous-time switched LTI systems. 
These conditions can be verified in polynomial-time, more precisely in $O((m(n+p))^2)$, where $n$ is the number of state variables, $p$ is the number of unknown inputs, and $m$ is the number of modes. 
Additionally, accounting for feed-forward scenarios, we addressed the minimum sensor placement for these systems designing an algorithm with a computational complexity of $O((m(n+p)+\alpha)^{2.373})$, where $\alpha$ is the number of target strongly connected components. 

Finally, we examined various assumptions on both the system and unknown inputs (latent space) dynamics. These assumptions translated into imposing additional structure on the  problem, which allowed for finding solutions in a more efficient manner (i.e., with lower computational complexity). These algorithms are suitable to handle large-scale systems. 

\section*{Appendix}
{\small 
\noindent\textbf{Proof of Theorem~\ref{th:struct_obs}:} 
The continuous-time switched LTI system with unknown inputs described in~\eqref{eq:switched_lti_a}, \eqref{eq:switched_lti_b}, and \eqref{eq:switched_lti_cprime} may be re-written as the following augmented continuous-time switched LTI system where the new augmented system is $x'={[d^{\intercal}\,\, x^{\intercal}]}^\intercal$, 
\begin{equation}\label{eq:switched_augm}
   \dot{x}'(t)  = \underbrace{\left[\begin{smallmatrix}
    Q_{\sigma(t)} & \mathbf{0} \\
    F_{\sigma(t)} & A_{\sigma(t)} \end{smallmatrix}\right]}_{A_{\sigma(t)}'}x'(t) \,\text{ and }\, 
    y'(t)  = \underbrace{\left[\begin{smallmatrix} D & C\end{smallmatrix}\right]}_{C_{\sigma(t)}'}x'(t).
\end{equation} 
Moreover, let $\mathbb M=\{1,\ldots,m\}$ be the ordered finite set of modes where the function $\sigma(t)$ is constant. Then, we have that 
$
A_k'= 
\left[\begin{smallmatrix}
     Q_k   &  \mathbf{0}   \\
     F_k &  A_k \\
\end{smallmatrix}\right]
\,\text{ and }\,
{C}' = 
\left[\begin{smallmatrix}
     D & C 
\end{smallmatrix}\right].
$ 
Therefore, when the system in~\eqref{eq:switched_lti_a}, \eqref{eq:switched_lti_b}, and \eqref{eq:switched_lti_cprime} is structurally state and input observable, it is equivalent to when the system in~\eqref{eq:switched_augm} is structurally state observable.
Interestingly, despite the fact that observability and controllability are not dual in general for switched LTI systems, in Theorem~4 in \cite{meng2006observability}, the authors showed that switched LTI systems are dual in the case of circulatory switching (see Definition~3 in \cite{meng2006observability}). From Remark~3 in~\cite{liu2013structural}, it readily follows that, in the context of structural switched LTI systems, the order of the switches does not play a role in attaining structural controllability. Therefore, in particular, it follows that structural controllability can be attained in the case of circulatory switching. As such, we can leverage Theorem~4 in \cite{meng2006observability} and invoke duality between structural controllability and structural (state) observability. Hence, by Theorem~3 of \cite{pequito2017structural}, the system in~\eqref{eq:switched_augm} is structurally observable whenever the conditions $(i)$ and $(ii)$ hold.

\noindent\textbf{Proof of Corollary~\ref{th:struct_obs_sol}:} First, we construct the augmented system~\eqref{eq:switched_augm} from the original system~\eqref{eq:switched_lti}. 
Second, we need to ensure that the conditions in Theorem~\ref{th:struct_obs} are satisfied. 
When the digraph $\mathcal G\left(\bigvee_{k=1}^m\bar A_k',
\bar C'\right)$ has no non-accessible output vertices, it is equivalent to the existence of an edge from a state variable in each target-SCC $\mathcal{G} \left(\bigvee_{k=1}^m\bar A_k'\right)$ to an output vertex of $\mathcal{ G}\left(\bigvee_{k=1}^m\bar A_k',\bar C'\right)$. Thus, condition $(i)$ is equivalent to condition $(i)$ of Theorem~\ref{th:struct_obs}. 
Subsequently, we recall the result from~\cite{commault2002characterization}, which states that for $\bar M\in\{0,\star\}^{n_1\times n_2}$, when the $\text{g-rank}(\bar M)=\min\{n_1,n_2\}$, it is equivalent to when there exists a maximum matching of $\mathcal B(\bar M)$ of size $\min\{n_1,n_2\}$. 
Hence, by the previous result, condition $(ii)$ is equivalent to condition $(ii)$ of Theorem~\ref{th:struct_obs}.

\noindent\textbf{Proof of Theorem~\ref{th:soundness_and_complexity}:} 
To address the problem $\mathcal P_1$, we augment the system in \eqref{eq:switched_lti_a} and \eqref{eq:switched_lti_b} to be written as in \eqref{eq:switched_augm} where $x'=[d^{\intercal}$ $x^{\intercal}]$. 
With this augmented system, Algorithm~\ref{alg:main} constructs three minimum sets of dedicated outputs that combine to satisfy the two conditions outlined in Theorem~\ref{th:struct_obs}, which guarantee structural state and input observability. Minimality of the combined sets is ensured as we use maximum matchings to build the three sets. 


Subsequently, we use Algorithm~\ref{alg:main} with the structural switched LTI system with $\mathbb M=\{1,\ldots,m\}$ modes described by the matrices $\{\bar A_1',\ldots,\bar A_m'\}$, where the matrices $\bar A_k'$ are defined as 
$\bar A_k'= 
\left[\begin{smallmatrix}
     \bar Q_k   &  \mathbf{0}   \\
     \bar F_k & \bar A_k 
\end{smallmatrix}\right]
, \forall k\in\mathbb{M}.$ 

First, we observe that $\mathcal J'$ comprises a minimum set of dedicated outputs, which maximizes the \\ $\text{g-rank}([\bar A_1';\ldots;\bar A_m';\mathbb I_{(n+p)}^{\mathcal J'}])$, where  $\mathbb I^{\mathcal J'}_{(n+p)}$ is a diagonal matrix whose entries in $\mathcal{J}'$ are nonzero. Concatenating  $[\bar A_1';\ldots;\bar A_m']$ with $\mathbb I^{\mathcal J'}_{(n+p)}$ increases the generic rank by $|\mathcal J'|$ and produces dedicated outputs assigned to state variables in distinct target-SCCs. 
In fact, $\mathcal B([\bar A_1';\ldots;\bar A_m'])$ yields a MWMM $\mathcal M$ with weight 0 and size $|\mathcal M|$. 
Hence, by the result from~\cite{commault2002characterization} used in the proof of Corollary~\ref{th:struct_obs_sol}, it follows that $\text{g-rank}([\bar A_1';\ldots;\bar A_m';\mathbb I^{\mathcal J'}_{(n+p)}])=|\mathcal M|$. 

Next, a MWMM $\mathcal M'$ of $\mathcal B([\bar A_1';\ldots;\bar A_m';\bar T^{\intercal}])$ has size $|\mathcal M'|$. 
This corresponds to an increase in \\ $\text{g-rank}([\bar A_1';\ldots;\bar A_m';\mathbb I^{(\mathcal J'\cup \mathcal J'')}_{(n+p)}])$ from \\ \mbox{$\text{g-rank}([\bar A_1';\ldots;\bar A_m';\mathbb I^{\mathcal J'}_{(n+p)}])$} of $|\mathcal M'|-|\mathcal M|$.  
Observe that, by the construction of the matrix $\bar T$, we have that $\mathbb I^{\mathcal J''}_{(n+p)}$ corresponds to dedicated outputs assigned to state variables in distinct target-SCCs. 
This means that $|\mathcal J''|$ target-SCCs will have outgoing edges to different outputs of the system digraph. 
This is necessary to satisfy condition $(i)$ of Theorem~\ref{th:struct_obs} but may not be sufficient. 

Therefore, we have to finally consider a third set, $\mathcal J'''$, to ensure that condition $(i)$ is fulfilled. 
In other words, there might still be target-SCCs that are not accounted for by state variables indexed in $\mathcal J'\cup\mathcal J''$, which we account for in $\mathcal J'''$. 

By minimizing the number of additional dedicated outputs $\mathbb I_{(n+p)}^{\mathcal J''}$, in step~8, we satisfy condition $(ii)$ in Theorem \ref{th:struct_obs} since \mbox{$\text{g-rank}([\bar A_1';\ldots;\bar A_m';\mathbb I^{(\mathcal J'\cup\mathcal J'')}_{(n+p)}]) = n+p$}. Additionally, the set $\mathcal J'''$ of minimum extra dedicated outputs, found in step~9, ensures that there are not state vertices that do not access at least one output vertex in $\mathcal G\left(\bigvee_{k=1}^m\bar A_k',\,\mathbb I^{\mathcal J}_{(n+p)}\right)$, where $\mathcal J=\mathcal J'\cup\mathcal J''\cup\mathcal J'''$, thereby fulfilling condition (i) of Theorem~\ref{th:struct_obs}. 
Notice that $\mathbb I^{\mathcal J''}_{(n+p)}$ are not
 assigned to previously assigned target-SCCs, as
 they would have been considered in $\mathbb I^{\mathcal J'}_{(n+p)}$. 
 
 Consequently, by the construction, setting $\mathcal J=\mathcal J'\cup\mathcal J''\cup\mathcal J'''$ in step~10 yields a solution $\mathbb I^{\mathcal J}_{(n+p)}$ that is minimal, ensuring both conditions of Corollary~\ref{th:struct_obs_sol}. 
 Notice that the produced solution easily translates to the original problem $\mathcal P_1$  solution by setting the originals $\bar C=\mathbb I_n^{\mathcal J_x}$ and $\bar D=\mathbb I_p^{\mathcal J_d}$, where $\mathcal J_x =\{i\in\mathcal J\,:\,i > p\}$ and $\mathcal J_d =\{i\in\mathcal J\,:\,i \leq p\}$.
 
 The computational complexity of Algorithm~\ref{alg:main} comes from the step with the highest computational cost (step~6) since the remaining steps of the algorithm have lower complexity. 
The computational complexity of step~6 can be solved by resorting to the Hungarian algorithm \cite{kuhn1955hungarian} that finds a MWMM of $\mathcal B([\bar A_1';\ldots;\bar A_m';\bar T])$ in $O(\max\{|\mathcal V_r|,|\mathcal V_c|\}^\varsigma )$, where $\mathcal V_r$ and $\mathcal V_c$ are defined in step~4 and $\varsigma<2.373$ is the exponent of the best known computational complexity of performing the product of two square matrices. 
Since $|\mathcal V_c|\leq|\mathcal V_r|$, this results in a computational cost of $O(|\mathcal V_r|^\varsigma)=O((m(n+p)+\alpha)^\varsigma)$.

\noindent\textbf{Proof of Theorem~\ref{th:soundness_and_complexity_class1}:} 
To prove the soundness of Algorithm~\ref{alg:class1}, we need to verify the two conditions in Corollary~\ref{th:struct_obs_sol}. In step~9, the set $\mathcal{J}'$ assigns an output to the nodes connected to the virtual target node $t$ that are associated with the obtained disjoint paths that start from the virtual source node and end in virtual target node. 
Subsequently, in step~10, $\mathcal J''$ assigns an output to a state variable of each target-SCC that is not contemplated in the set $\mathcal J'$. 
Therefore, $\mathcal J=\mathcal J'\cup\mathcal J''$ ensures condition (i) of Corollary~\ref{th:struct_obs_sol}. 
Next, we set $\bar C'=\mathbb I^{\mathcal J}$. 
    
Now, we can see that $\mathcal B\left( [\bar A_1'; \ldots; \bar A_m';\bar C' ]\right)$ has a maximum matching of size $n+p$, so condition (ii) of Corollary~\ref{th:struct_obs_sol} is satisfied, corresponding to a decomposition of the system's digraph into paths and cycles that span the digraph, whose paths end in output vertices. This property holds because the system's digraph $\mathcal{G}(\bigvee_{k=1}^{m}\bar A_{k}')$ is spanned by the paths obtained in step~8, together with the edges that connect the end of the paths to the respective outputs and the cycles (self-loops) from the state variables that do not belong to the paths. Thus, Algorithm~\ref{alg:class1} is sound. 
%

The computational complexity of Algorithm~\ref{alg:class1} comes from the step with the highest computational cost (step 7). The computational complexity of step 7 can be solved by finding the vertex disjoint paths, which is linear in the number of edges \cite{kawarabayashi2012disjoint}, and hence, quadratic in the number of vertices. 

\noindent\textbf{Proof of Theorem~\ref{th:soundness_and_complexity_class2}:} 
Since the system is spanned by cycles as a result of the nodal dynamics, then the system is structurally state and input observable if and only if all of the states and inputs are accessible \cite{ramos2015analysis}. Hence, by guaranteeing that at least one state in each target-SCCs is measured, we ensure that the system is~accessible.

The complexity follows from the fact that each step is linear in time, except for finding the target-SCCs where the computational complexity is linear in the number of nodes plus edges resulting from Tarjan's strongly connected components algorithm~\cite{cormen2009introduction}, which in the worst case is $O((n+p)^2)$.
}

%

{\small
\bibliographystyle{plain}
\bibliography{autosam} 
}
\end{document}